\documentclass[twocolumn]{aastex631}

\usepackage{graphicx}
\usepackage{amssymb}

\received{}
\revised{}
\accepted{}
\submitjournal{AAS}

\shorttitle{Eclipsing Binaries AB Cas and OO Dra}
\shortauthors{Kim}

\begin{document}
\title{Asteroseismic Determination of Stellar Rotation: On Synchronization in the Close Eclipsing Binaries AB Cas and OO Dra}
\correspondingauthor{Seung-Lee Kim}
\email{slkim@kasi.re.kr}
\author[0000-0003-0562-5643]{Seung-Lee Kim}
\affil{Korea Astronomy and Space Science Institute, Daejeon 34055, Republic of Korea}

\begin{abstract}
A star's rotation rate is difficult to estimate without surface inhomogeneities such as dark or bright spots.
This paper presents asteroseismic results to determine the rotation rates of $\delta$ Sct-type pulsating primary stars in two eclipsing binary systems, AB Cas and OO Dra.
After removing the binarity-induced light variations from the archival TESS data and carefully examining the combination frequencies, 
we identified 12 independent frequencies for AB Cas and 11 frequencies for OO Dra, with amplitudes higher than $\sim$0.3 mmag, as $\delta$ Sct-type pulsation frequencies excited in each primary star.
The theoretical frequencies for seismic analysis were obtained by fully considering the rotation effects.
Grid fitting for various stellar properties, such as mass, radius, metallicity, and rotation rate, yielded the best solution for which theoretical frequencies and stellar parameters agreed well with the observations.
The rotation rate of the AB Cas primary was tightly constrained to 0.81 $\pm$ 0.01 day$^{-1}$ ($f_{\rm rot} / f_{\rm orb}$ = 1.11$^{+0.01}_{-0.02}$), which is slightly faster than the synchronized rotation.
In contrast, the rotation rate of 0.63 $\pm$ 0.01 day$^{-1}$ for the OO Dra primary is lower than the synchronous value of approximately 0.81 day$^{-1}$.
Subsynchronous rotation is uncommon in short-period binaries, and its physical mechanism is not yet well understood.
Our results show that asteroseismology can be used to precisely measure the rotation rate of fast-rotating $\delta$ Sct stars and thus provide a valuable constraint on rotation-orbit synchronization in close binary systems.
\end{abstract}

\section{Introduction \label{sec_intro}}
Rotation is an important physical parameter that affects the stellar structure and evolution \citep{maeder2000}.
A representative observable parameter for the stellar rotation is the projected rotational velocity $v \sin i$, measured from the Doppler broadening of the spectral lines.
It is a minimum estimate of a star's rotational velocity, whose actual value cannot be obtained without further information regarding the inclination angle ($i$) between the stellar rotation axis and line of sight.
However, the rotation axes of stars are distributed randomly in space, and it is difficult to determine the inclination angles for most stars, 
with some exceptions such as eclipsing binaries with the Rossiter-McLaughlin effect \citep{albrecht2007,albrecht2009}, solar-like oscillators \citep{gizon2003},
and rapidly rotating stars with a strong gravity-darkening effect \citep{monnier2007,zhao2009}.

Another observable parameter is the rotation rate $f_{\rm rot}$, which is the reciprocal of the rotation period.
If there are surface inhomogeneities, such as dark spots and bright faculae, the star's brightness varies along with its rotation \citep{strassmeier2009,lurie2017}.
Thus, the rotation rate can be accurately estimated from the rotational light variation, independently of the inclination angle.
These kinds of photometric variabilities are observed in hot stars with late-B or A spectral types \citep{balona2013,sikora2020} as well as in cool stars \citep{reinhold2020},
owing to the high-precision data from the space missions such as Kepler and TESS \citep[Transiting Exoplanet Survey Satellite,][]{ricker2015}.

The rotation rate can also be derived from asteroseismic analysis of pulsating stars if rotationally split multiplets of nonradial pulsation frequencies are detected;
stellar rotation makes the nonradial ($\ell \ne 0$) pulsation modes split into a series of the azimuthal order ($m$) from $-\ell$ to $+\ell$.
This approach produced unique results, such as measuring the rotation rate of the deep core as well as the surface in A-F main-sequence stars \citep{kurtz2014,saio2015,li2020}
and detection of latitudinal differential rotation in sun-like stars \citep{benomar2018}.
To date, a majority of these results were limited to slowly rotating stars for which the structural deformation effect caused by the centrifugal force was negligible.

$\delta$ Sct-type pulsating variables are mostly in the main-sequence phase of evolution, with spectral types ranging from early A to early F.
Many rotate at high speed with projected surface velocities around 100 km s$^{-1}$ or higher up to 300  km s$^{-1}$, 
whereas high-amplitude $\delta$ Sct stars (HADS) with single or double radial modes rotate slowly with $v \sin i \le$ 30 km s$^{-1}$ \citep{breger2000}.
$\delta$ Sct stars with multiple frequencies have long been considered attractive objects in asteroseismology \citep{brown1994}.
However, studies have not yet yielded fruitful results, mainly due to the difficulties of mode identification \citep{goupil2005,handler2013,kurtz2022}.
The fast rotation of $\delta$ Sct stars destroys the equidistance of rotational splitting \citep{goupil2000}, hindering mode identification based on regular frequency spacing \citep{paparo2016}.
Recently, several seismic interpretations of fast-rotating $\delta$ Sct stars have been performed using realistic two-dimensional stellar models;
examples include \citet{zwintz2019} for $\beta$ Pic and \citet{bouchaud2020} for Altair.

In our previous paper \citep{kim2021}, we introduced a new approach for the seismic analysis of fast-rotating $\delta$ Sct stars.
It is now called the {\it Interpolated Complete-Calculation} (ICC) approach, which estimates the rotational shifts of pulsation frequencies
by interpolating the complete calculations of \citet{reese2006b} that fully include the rotation effects based on a two-dimensional polytropic model.
In our previous study, we applied the ICC approach to the $\delta$ Sct-type pulsating primary in the EL CVn-type eclipsing binary system 1SWASP J024743.37-251549.2 (hereafter referred to as J0247-25)
and demonstrated that the theoretical values of pulsation frequencies, pulsation modes, and stellar parameters concurred well with the observations.
To our best knowledge, this was the first time that the polytropic model frequencies obtained from the complete calculations were used for the seismic analysis of $\delta$ Sct-type pulsators.

To validate the ICC approach, this study revisits two well-known objects, $\delta$ Sct-type pulsating primary stars in the double-lined eclipsing binary systems
AB Cas\footnote{The primary star of AB Cas is the first $\delta$ Sct-type pulsating component identified in Algol-type binary systems and is a member of the oscillating Eclipsing Algol (oEA).
The oEA stars, defined as mass-accreting pulsating components of A-F spectral type main sequence in semi-detached Algols, have evolved differently from the classical $\delta$ Sct stars \citep{mkrtichian2004,mkrtichian2022}.
Eclipsing binaries with various types of pulsating stars were reviewed by \citet{lampens2021}.} and OO Dra.
These two $\delta$ Sct primaries differ from classical $\delta$ Sct stars in evolutionary terms due to the mass transfer between the binary components \citep{hong2017,lee2018,mkrtichian2022};
the binary evolution is described in Section \ref{sec_syncCBS}.
They have an important advantage in the asteroseismic aspect that their physical properties, such as masses and radii, are well constrained from the photometric and spectroscopic observations.
The properties of the binary components, taken from \citet{hong2017} for AB Cas and \citet{lee2018} for OO Dra, are listed in Table \ref{tab_param}.

The asteroseismic analysis of these two objects was recently conducted.
\citet{miszuda2022} analyzed the primary star of AB Cas by fixing the rotation rate as the synchronous value to the orbit.
The seismic result by \citet{chen2021a} showed that the OO Dra primary rotated slightly faster than the synchronous value, which is in contrast to the spectroscopic observations by \citet{lee2018}.
Both seismic studies estimated the rotation effects on pulsation by applying the perturbative approach valid for slow rotators.
According to the numerical results of \citet{reese2006b}, using complete calculations of the rotation effects is required to correctly interpret fast-rotating pulsators with $v > 50$ km s$^{-1}$.

The previous studies by \citet{miszuda2022} and \citet{chen2021a} performed the frequency analysis using the archival TESS data.
After subtracting the binarity-induced light variations, they analyzed the residual data for all orbital phases.
If the primary star is considerably eclipsed by the secondary component, brightness variations due to the pulsation of the primary are distorted during the eclipse.
The eclipsed area of the AB Cas primary is not negligible to be approximately 84\% at maximum, and the OO Dra primary is eclipsed up to approximately 33\%.
Therefore, it is better to exclude the primary eclipse data when conducting the frequency analysis.
Furthermore, the TESS data of OO Dra were doubled by adding two Sectors after \citet{chen2021a}.
Since the data selection differing from the previous studies can affect the frequency analysis, we decided to reanalyze the TESS data of two binaries in this study.
Data reduction and frequency analysis are presented in Section \ref{sec_obsdata} and Section \ref{sec_frequency}, respectively.
Our seismic analysis applying the ICC approach is described in Section \ref{sec_seismology}.
Finally, in Section \ref{sec_discussion}, we discuss the seismic results and the rotation-orbit synchronization in close binary systems.

\begin{deluxetable}{rcc}
\tablecaption{Physical Properties} \label{tab_param}
\tablehead{ \colhead{Parameter} & \colhead{AB Cas} & \colhead{OO Dra} } 
\startdata
$M_1$ ($M_\sun$)    & 2.01 $\pm$ 0.02  & 2.03 $\pm$ 0.06  \\
$M_2$ ($M_\sun$)    & 0.37 $\pm$ 0.02  & 0.19 $\pm$ 0.01 \\
$R_1$ ($R_\sun$)     & 1.84 $\pm$ 0.02  & 2.08 $\pm$ 0.03 \\
$R_2$ ($R_\sun$)     & 1.69 $\pm$ 0.03  & 1.20 $\pm$ 0.02 \\
$T_1$ (K)                  & 8080 $\pm$ 170  & 8260 $\pm$ 210 \\
$T_2$ (K)                  & 4925 $\pm$ 150  & 6268 $\pm$ 150 \\
$v_1 \sin i$ (km s$^{-1}$)            & 73 $\pm$ 3  & 72 $\pm$ 5 \\
$v_{1,\rm sync}$ (km s$^{-1}$)   & 68 $\pm$ 1  & 85 $\pm$ 1 \\
$r_{1,\rm side}$      & 0.2674 $\pm$ 0.0003 & 0.3303 $\pm$ 0.0005 \\
$r_{1,\rm back}$     & 0.2686 $\pm$ 0.0003 & 0.3317 $\pm$ 0.0005 \\
$r_{1,\rm volume}$ & 0.2669 $\pm$ 0.0003 & 0.3287 $\pm$ 0.0005 \\
$a$ ($R_\sun$)         & 6.91 $\pm$ 0.06  & 6.33 $\pm$ 0.06 \\
$i$ (deg)                   & 89.9 $\pm$ 0.2    & 85.3 $\pm$ 0.1 \\
$\gamma$ (km s$^{-1}$)           & $-$67.5 $\pm$ 1.2  & $-$52.8 $\pm$ 0.6 \\
\enddata
\tablenotetext{}{{\bf Note.} The subscripts 1 and 2 indicate the primary and secondary components, respectively. $v_{\rm sync}$ is the predicted velocity at the equator for synchronous rotation.
Data were taken from \citet{hong2017} for AB Cas and \citet{lee2018} for OO Dra.}
\end{deluxetable}

\section{Data Reduction \label{sec_obsdata}}
\subsection{TESS Data \label{sec_tess}}
We used two-minute cadence data from NASA's all-sky survey mission, TESS.
AB Cas was observed with TESS during Sectors 18, 19, and 25 from November 3, 2019, to June 8, 2020 (from BJD 2,458,790.66123 to 2,459,009.30165).
OO Dra was observed in Sectors 20 and 21 from December 24, 2019, to February 18, 2020 (from BJD 2,458,842.50623 to 2,458,897.78267),
and revisited in Sectors 40 and 41 from June 25 to August 20, 2021 (from BJD 2,459,390.65116 to 2,459,446.57606).

Data were downloaded from the Mikulski Archive for Space Telescopes (MAST, https://mast.stsci.edu,
\dataset[10.17909/t9-nmc8-f686]{https://doi.org/10.17909/t9-nmc8-f686}).
From the data files, we extracted the observation times (BJD $-$ 2,457,000) and the Simple Aperture Photometry (SAP) fluxes in electrons per second;
the Pre-search Data Conditioning Simple Aperture Photometry (PDCSAP) fluxes were not adopted in this study
because the PDCSAP fluxes in Sector 18 for AB Cas showed irregular light variations seriously differing from the SAP fluxes and other Sectors' data,
and because the correction process of the PDCSAP fluxes can erase true variability in stars \citep{raiteri2021}.
The SAP flux was converted into a TESS magnitude using a simple flux-to-magnitude equation, $mag = -2.5 \log_{10} (flux) + Z_{\rm T}$, 
where the zero point $Z_{\rm T}$ was set to 20.44 mag, following \citet{fausnaugh2021}.
Anomalous data with a nonzero "QUALITY" flag were excluded.

\begin{figure*}
\center\includegraphics[scale=0.58]{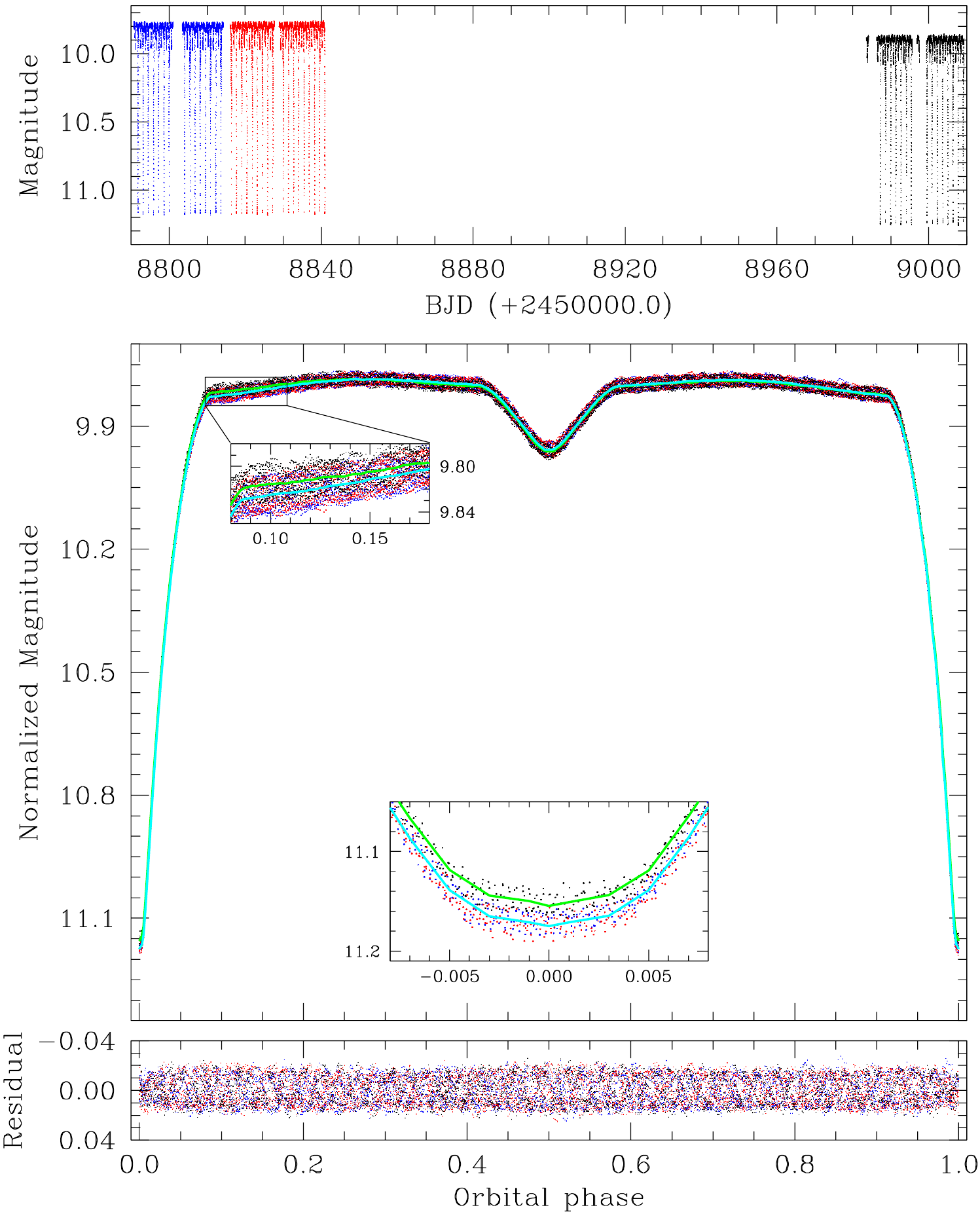}\hskip 3mm\includegraphics[scale=0.58]{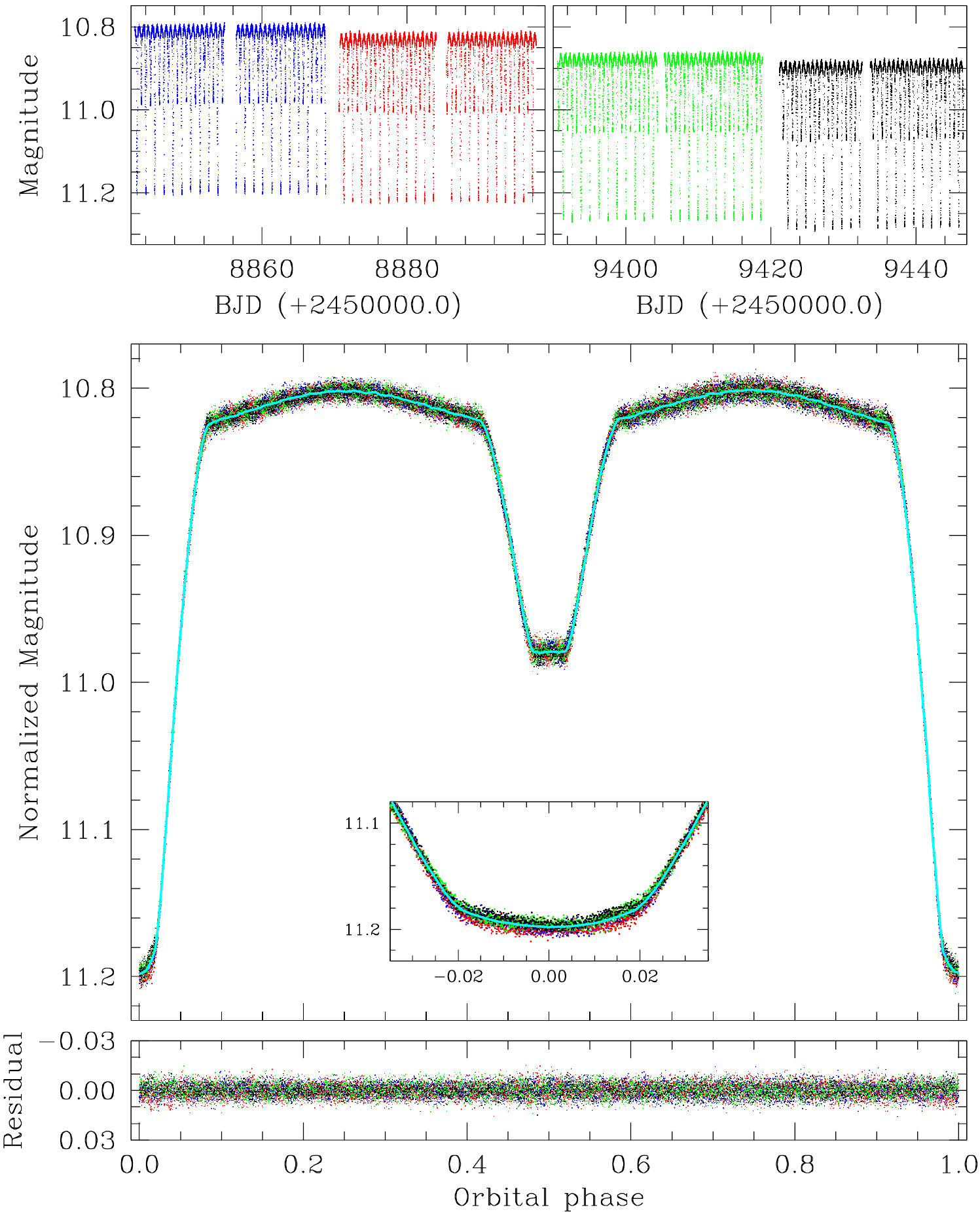}
\caption{TESS light curves of AB Cas (left) and OO Dra (right). Small dots represent the data, with different colors for each Sector.
Phase diagrams are displayed in the middle after normalizing the magnitude differences between Sectors (also between the TESS orbits).
Solid curves are mean magnitudes calculated with a 0.002 phase interval; for AB Cas, the mean curve of Sector 25 (green line) is slightly different from that of the other two consecutive Sectors (cyan line),
particularly during and after the primary eclipse, and therefore calculated separately. The insert panels are zoom-in views.
The residuals from the mean curves are presented at the bottom, showing relatively large dispersions in AB Cas caused by the dominant pulsation frequency with a high amplitude. \label{fig_lc}}
\end{figure*}

\subsection{Light Curves of Eclipsing Binaries \label{sec_lc}}
We derived 48 and 81 new times of primary-minimum light from the TESS data for AB Cas and OO Dra, respectively.
These times were linearly correlated with the number of orbital cycles, yielding orbital periods of 1.3668840(8) days for AB Cas and 1.2383814(1) days for OO Dra.
Our estimates are nearly identical to the recent results, that is, 1.3668810(6) days for AB Cas obtained by \citet{miszuda2022} from the Fourier analysis of the TESS data,
and 1.2383813(50) days for OO Dra obtained by \citet{lee2018}, also adopted by  \citet{chen2021a}, from simultaneous analysis of ground-based photometric and spectroscopic data.
The new epochs of the primary minima were BJD 2,458,840.98503(5) for AB Cas and 2,458,897.28346(2) for OO Dra.
The values in parentheses are the errors in the last digit.

During the early stages of data reduction, we found that the phase-folded light curves were well-defined for each Sector but showed a conspicuous discrepancy in magnitude between Sectors.
The top panels of Figure \ref{fig_lc} show this discrepancy of up to approximately 0.1 mag, indicating that the zero point differs from Sector to Sector \citep{raiteri2021}.
The data of each Sector are divided into two parts separated by an observational gap, and the mean magnitude of the first half is slightly different from that of the second half;
the TESS observations are paused for data downlink in the middle of each Sector, at every perigee pass of the elliptical orbit with a period of 13.7 days, and
the spacecraft reorientation for data downlink changes the camera temperatures, leading to a shift in the bias level \citep{vanderspek2018}.
Therefore, the magnitudes were normalized to the data of the first orbit by comparing the average values of the out-of-eclipse phase.
For AB Cas, averaging was conducted after subtracting light variations of the dominant pulsation frequency with a high amplitude (see Table \ref{tab_ABobsfreq}).
Light curves with normalized magnitudes are displayed in the middle of the figure, where the orbital phases were calculated using our new orbital periods and minimum epochs.

To remove binarity-induced light variations (eclipse, ellipsoidal effect, and reflection effect, etc.) and to obtain residual data suitable for asteroseismic analysis,
the mean magnitudes were calculated with a 0.002 phase interval (500 bins spanning the orbital period).
The mean curves do not show any noticeable wiggles, as presented in the middle panels of Figure \ref{fig_lc}, because each bin contains sufficient data points.
The number of bins was tested for several cases and did not affect frequency analysis \citep{chen2021a}.
For AB Cas, the mean curve of Sector 25 was slightly different from that of the other two consecutive Sectors, particularly during and after the primary eclipse.
This unusual feature appears to be caused by variations in the transfer mass flow or starspots on the surface of the binary components.
The bottom panels of Figure \ref{fig_lc} show the residuals without any peculiarity; this study and that of \citet{chen2021a} for OO Dra derived the residual data after subtracting the mean curves,
but in the previous research of \citet{miszuda2022} for AB Cas, the residuals were obtained from the model light curve and showed some systematic variations owing to the incompleteness of binary light-curve modeling.

Both binaries show transit-like primary minima with limb-darkening curvatures as presented in the inset panels of Figure \ref{fig_lc}.
These eclipses confirm the high orbital inclination close to 90 deg, 
which provides an important constraint on the visibility of nonradial pulsation modes (see Section \ref{theo_freq}).

\begin{figure*}
\center\includegraphics[scale=0.60]{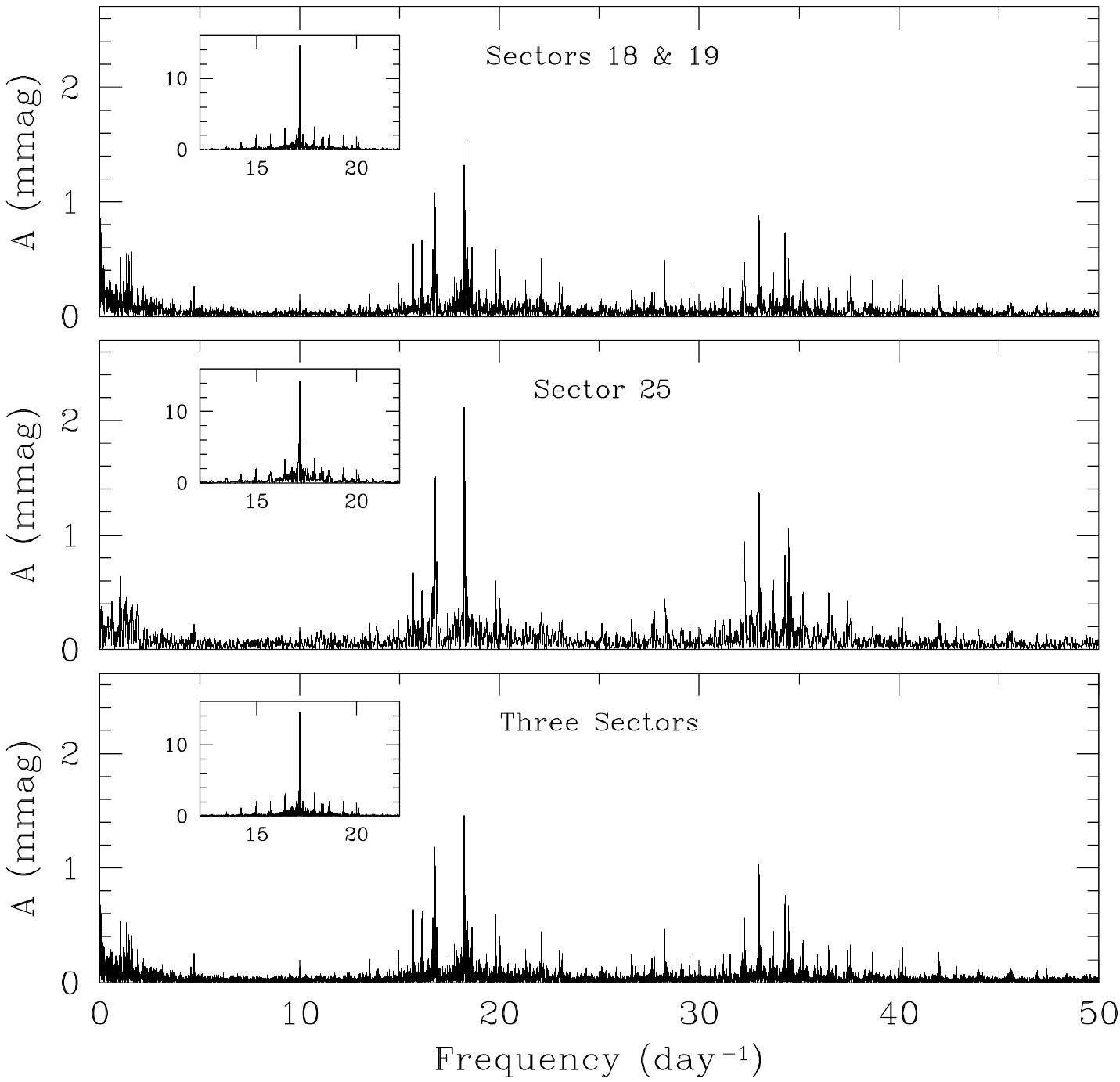}\hskip 3mm\includegraphics[scale=0.60]{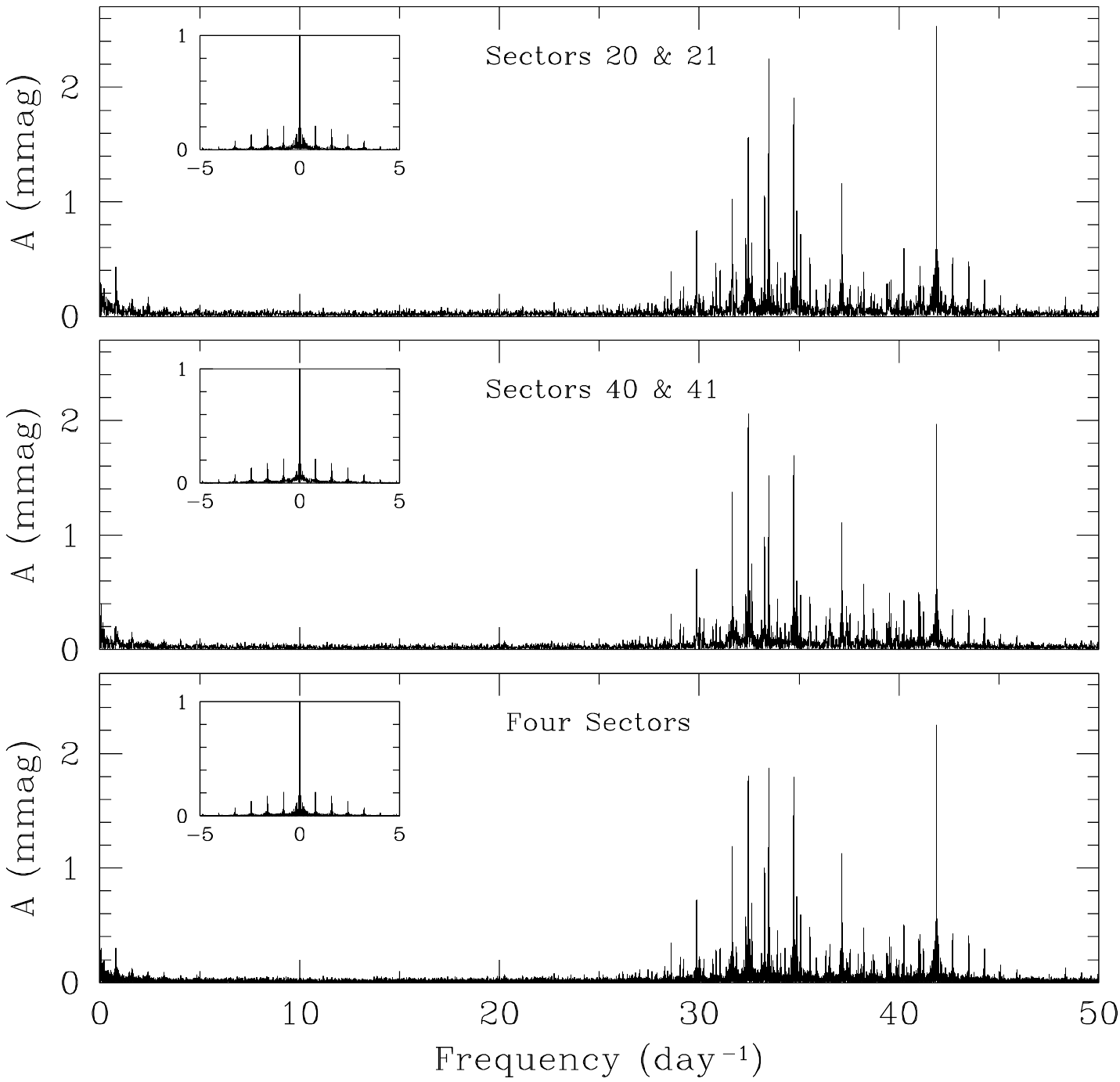}
\caption{Fourier amplitude spectra of AB Cas (left) and OO Dra (right), using the residuals except for the primary eclipse phase.
The inset panels on the left show the first frequency of AB Cas with a dominant amplitude, mimicking the window spectra of OO Dra on the right. 
The left spectra were obtained after removing the first frequency of AB Cas. \label{fig_fourier}}
\end{figure*}

\section{Frequency Analysis \label{sec_frequency}}
\subsection{Detection of Multiple Frequencies}
A multiple frequency analysis was conducted by applying the discrete Fourier transform and least-squares fitting process \citep{kim2010}.
Figure \ref{fig_fourier} shows the Fourier amplitude spectra of the residuals after subtracting the mean curves from the observed data.
The residuals during the primary eclipse were not used for the analysis 
because the $\delta$ Sct-type primary component was considerably eclipsed by the secondary star.
Thus, several weak sidelobes equally spaced by the orbital frequency ($f_{\rm orb}$) can be observed in the window spectra.

More than 100 frequencies for each target were detected from the consecutive prewhitening process by applying the empirical criterion of a signal-to-noise amplitude ratio S/N $\ge$ 4.0 \citep{breger1993}.
Owing to the high-precision TESS data, the frequencies with a very low amplitude, down to 0.07 mmag, passed this criterion.
Most of the low-amplitude frequencies seem to be identified as combination frequencies \citep[for example,][]{breger2011,liakos2018},
and some can be $\delta$ Sct-type pulsation frequencies with high-degree modes of $\ell \ge 4$ or $\ell + |m| = odd$ modes.
The former frequencies are not used for asteroseismic analysis, and the latter ones are not covered in our seismic analysis (see Section \ref{theo_freq}).
Consequently, only frequencies with amplitudes higher than $\sim$0.3 mmag were considered in this study, as listed in Appendix Tables \ref{tab_ABobsfreq} and \ref{tab_OOobsfreq}.

Our results for frequencies greater than 2.0 day$^{-1}$ are mostly identical to the previous results of \citet{miszuda2022} for AB Cas and \citet{chen2021a} for OO Dra.
However, as separately presented at the bottom of the Appendix tables, some frequencies are detected in only one result, and some have amplitudes significantly different between the previous result and ours.
In particular, many combinations of the dominant pulsation frequency with the orbital frequency, $f_1 + N\,f_{\rm orb}$,
found by \citet{miszuda2022} for AB Cas were not detected or showed to be very low amplitude in our study.
It is probably because the previous analysis of AB Cas included the primary eclipse data of which pulsation amplitudes considerably decreased due to the eclipse, as shown in Figure 3 of \citet{miszuda2022}.
For OO Dra, four high frequencies are different between the results by \citet{chen2021a} and ours.
Two unresolved frequencies, $f_{16} (=f_2)$ and $f_{24} (=f_1)$, were not detected in the previous analysis due to the low resolution by the short-run data of approximately 58 days.
The amplitudes of two orbital harmonic frequencies, $f_6 (=46\,f_{\rm orb})$ and $f_7 (=37\,f_{\rm orb})$, may be reduced by some wiggles seen in the mean curve of \citet{chen2021a}.

In contrast, most frequencies less than 2.0 day$^{-1}$ were not previously detected.
Some of these low frequencies are combinations, and others are likely instrumental artifacts that are not sufficiently detrended.
Although low frequencies can originate from A-type main-sequence stars with $\gamma$ Dor-type gravity-mode pulsations, it is difficult to anticipate the $\gamma$ Dor-type pulsation frequencies in this study
because the AB Cas primary and OO Dra primary are hotter than the blue edge of the theoretical $\gamma$ Dor instability strip \citep{xiong2016}.
Since we focus on $\delta$ Sct-type pressure-mode pulsations with frequencies greater than 4.0 day$^{-1}$, the low frequencies are not further examined.

\subsection{Identification of Independent Frequencies \label{indep_freq}}
The proper selection of independent pulsation frequencies is imperative for successful asteroseismic analysis.
A careful examination was performed to select and discard the combination frequencies $f_c = M\,f_i + L\,f_j$, where $M$ and $L$ are integers.
In general, the combination frequencies $f_c$ have amplitudes much lower than the parent (or independent) frequencies $f_i$ and $f_j$ \citep{breger2008,breger2011,lv2022},
which is more severe in higher-order terms.
\cite{papics2012} recommended that one should try to find the higher-order combinations only when the presence of second-order terms is proven.
The frequencies considered in this study have a narrow range of amplitudes from $\sim$0.3 mmag to $\sim$2 mmag, except for $f_1$ of AB Cas,
and therefore we analyzed only the second-order terms by setting ($|M|+ |L|$) = 2.

For binary systems, orbital harmonics ($N\,f_{\rm orb}$) and combinations with orbital harmonics ($M\,f_i + K\,f_{\rm orb}$) were detected
with similar amplitudes as the independent frequencies \citep[for example,][]{wang2019,kim2021}.
Combinations with orbital harmonics can occur by several factors, such as the variation in the light contribution with the orbital phase \citep{kim2010,miszuda2022}.
We examined the orbital harmonics for the frequency range of 0.0--50.0 day$^{-1}$ (for example, $N \le 68$ for AB Cas) and combinations with orbital harmonics up to $|K| \le 15$.

The frequency resolution criterion of 1.5 times the Rayleigh limit \citep{loumos1978} was used for these selection processes, 
that is, 1.5/$\Delta$T = 0.00686 day$^{-1}$ for AB Cas and 0.027 day$^{-1}$ for OO Dra,
identical to the previous studies, where $\Delta$T is the total span of observations (for example, $\Delta$T = 218.64 days for AB Cas).
Although we analyzed a much larger dataset of OO Dra compared to \citet{chen2021a}, we adopted the duration of two consecutive Sectors as $\Delta$T,
instead of the total span of observations, considering the long observational gaps of approximately 493 days;
\citet{breger2002} commented that the frequency resolution of two months observed one year apart corresponds to that of a single month.

Consequently, 12 frequencies for AB Cas and 11 frequencies for OO Dra were identified as independent frequencies that are $\delta$ Sct-type pulsation frequencies excited in the primary components,
as represented by the bold characters in the Appendix Tables \ref{tab_ABobsfreq} and  \ref{tab_OOobsfreq}.
Among them, 10 frequencies of AB Cas and seven for OO Dra are identical to those previously identified by \citet{miszuda2022} and \citet{chen2021a},
respectively.\footnote{\citet{miszuda2022} identified a total of 17 independent frequencies for AB Cas, of which 10 are identical to ours, and one is a low frequency of less than 1.0 day$^{-1}$.
The other six frequencies have amplitudes less than 0.3 mmag and, therefore, are not considered in this study. All seven independent frequencies identified by \citet{chen2021a} for OO Dra are identical to ours.}
Other frequencies were identified previously as high-order combination frequencies.
\citet{miszuda2022} identified two frequencies of AB Cas, $f_{15}$ = 32.2506 day$^{-1}$ and $f_{22}$ = 16.8874 day$^{-1}$, as fourth-order combinations, 
$2\,f_{\rm M22\_4} - 2\,f_{\rm M22\_32}$ and $3\,f_{\rm M22\_7} + f_{\rm M22\_19}$, respectively.
\citet{chen2021a} identified four frequencies of OO Dra, $f_{13}$ = 39.5191 day$^{-1}$, $f_{15}$ = 34.9254 day$^{-1}$, $f_{22}$ = 39.6160 day$^{-1}$, and $f_{23}$ = 37.5473 day$^{-1}$, 
as third-order combinations, $f_{\rm C21\_3} + f_{\rm C21\_11} - f_{\rm C21\_2}, f_{\rm C21\_3} + f_{\rm C21\_7} - f_{\rm C21\_4}, f_{\rm C21\_1} + f_{\rm C21\_7} - f_{\rm C21\_10}$,
and $2\,f_{\rm C21\_8} - f_{\rm C21\_7}$, in order.
These high-order combinations are probably a mere coincidence, considering that their amplitudes are not too low, and their second-order terms are not detected.\footnote{Since
it may be questionable whether these frequencies are independent (this study) or combinations (previous study),
we checked what would happen if the seismic analysis described in Section \ref{sec_seismology} was conducted, excluding these frequencies. 
We found that including or excluding these frequencies produces the same results for the rotation rates and, therefore, does not change the conclusion of this study.}

Three frequencies of AB Cas were identified as second-order combinations, $f_8 = 2\,f_1$, $f_{12} = f_2 - f_7$, and $f_{28} = 2\,f_3$ (Table \ref{tab_ABobsfreq}),
whereas no such combination was found for OO Dra (Table \ref{tab_OOobsfreq}).
The amplitude ratios between the combination ($f_c$) and parent ($f_i, f_j$) frequencies, defined as $A_{c} / (A_i A_j)$ with an amplitude unit of mmag,
were calculated to be 0.0031 for $f_8$, 0.63 for $f_{12}$, and 0.21 for $f_{28}$ of AB Cas.
The ratio for $f_8$ is in good agreement with known values of 0.0025--0.0092 \citep{breger2008,breger2011}, but those for $f_{12}$ and $f_{28}$ are much higher than the known values.
Therefore, it is doubtful whether the two frequencies $f_{12}$ and $f_{28}$ are real combinations.
We found the theoretical frequencies in the best solution (see Section \ref{sec_solution}), 18.372 day$^{-1}$ with mode ($n$ = 0, $\ell$ = 3, $m$ = +3) and 36.399 day$^{-1}$ with mode ($n$ = 4, $\ell$ = 2, $m$ = 0),
corresponding to $f_{12}$ = 18.258 day$^{-1}$ and $f_{28}$ =  36.493 day$^{-1}$, respectively.
Although $f_{12}$ and $f_{28}$ are possible pulsation frequencies excited in the primary star of AB Cas, these were not used for our seismic analysis.

\section{Asteroseismic Analysis \label{sec_seismology}}
The observed independent frequencies were compared with the theoretical frequencies.
The asteroseismic analysis was performed by applying the same approach as in \citet{kim2021}, and thus it is briefly described here.

\subsection{Theoretical Frequencies \label{theo_freq}}
Stellar evolutionary models were obtained using the Modules for Experiments in Stellar Astrophysics \citep[MESA,][]{paxton2011, paxton2013, paxton2015, paxton2018, paxton2019} software version 11701.
The evolution of a single star was applied, considering the results that the theoretical frequencies of a binary component are very close to those of a single star
with the same physical properties of mass, radius, and metallicity; the frequency differences between the binary and single-star models are approximately 0.01 day$^{-1}$ \citep{streamer2018, chen2021a}.
The initial helium abundance was set to $Y$ = 0.245 + 0.827\,$Z$ \citep{nsamba2021} as a function of the metallicity $Z$,
and the mixing length parameter was set to $\alpha_{\rm MLT}$ = 1.6.
For these stellar models without rotation, the theoretical pulsation frequencies were calculated under adiabatic conditions using the GYRE code \citep{townsend2013}.
The frequencies were scanned from a minimum of 10 day$^{-1}$ to a maximum of 100 day$^{-1}$ for the angular degree $\ell$ = 0--3 modes.
The other parameters are identical to those in \citet{kim2021} and are mostly left in the default setting.
Sample input files for MESA and GYRE calculations are available via Zenodo at \dataset[10.5281/zenodo.7641868]{https://doi.org/10.5281/zenodo.7641868}.

We added the nonrotating model frequencies from GYRE and their rotational shifts estimated from the two-dimensional numerical results by \citet{reese2006b}.
The results from complete calculations, fully including the rotation effects based on the centrifugally deformed polytropic model, are presented in \citet{reese2006a},
where a long list of pulsation frequencies with different rotation rates and pulsation modes ($n$ = 1--10, $\ell$ = 0--3, and $|m|$ = 0--3) is provided.
These values from the polytropic model were interpolated to derive frequency shifts for a realistic stellar model.\footnote{The three frequencies of AB Cas, $f_2$, $f_{20}$, and $f_{22}$,
were identified as nonradial modes with $n$ = 0 and $\ell$ = 2 or 3 (see Table \ref{tab_seismic} and Figure \ref{fig_echelle}).
Their dimensionless angular frequencies $\omega_0$ from GYRE were slightly lower than those of the $n$ = 1 mode from the polytropic model.
Thus, their rotational shifts were estimated from extrapolation.}

The two eclipsing binary systems AB Cas and OO Dra have high orbital inclinations, close to 90 deg (Table \ref{tab_param}).
In this particular case, the disc-integrated lights of the $\ell + |m| = odd$ modes exhibit very low amplitudes \citep{mkrtichian2018},
assuming that the rotation axis aligns with the orbital axis.
Therefore, only the $\ell + |m| = even$ mode frequencies were considered, reflecting the visibility of the nonradial pulsation modes.

\begin{deluxetable}{rcc}
\tablecaption{Astrometric and Kinematic Properties} \label{tab_kinematic}
\tablehead{ \colhead{Parameter} & \colhead{AB Cas} & \colhead{OO Dra} } 
\startdata
$\pi$ (mas)                                                 & 3.0323 $\pm$ 0.0151       & 1.4474 $\pm$ 0.0152 \\
$\mu_\alpha \cos \delta$ (mas yr$^{-1}$)  & $-$33.069 $\pm$ 0.012    & \,15.711 $\pm$ 0.017 \\
$\mu_\delta$ (mas yr$^{-1}$)                    & \,\,\, 19.733 $\pm$ 0.018   & $-$2.137 $\pm$ 0.018 \\
$U_{\rm LSR}$ (km s$^{-1}$)                    & \,99.2 $\pm$ 2.5                & \,80.3 $\pm$ 2.5 \\
$V_{\rm LSR}$ (km s$^{-1}$)                    & $-0.9$ $\pm$ 2.2             & $-4.8$ $\pm$ 2.0 \\
$W_{\rm LSR}$ (km s$^{-1}$)                   & \,\,\, 2.3 $\pm$ 1.7                  & $-8.8$ $\pm$ 1.7 \\
\enddata
\tablenotetext{}{{\bf Note.} Astrometric data of parallax ($\pi$) and proper motion ($\mu_\alpha \cos \delta$, $\mu_\delta$) were taken from the Gaia DR3 \citep{Gaia2016,Gaia2022}.
Galactic space-velocity components ($U_{\rm LSR}$, $V_{\rm LSR}$, $W_{\rm LSR}$) with respect to the local standard of rest (LSR) were calculated by the method of \citet{johnson1987}
and corrected for the solar motion of ($U_\sun$, $V_\sun$, $W_\sun$) = (9.58 $\pm$ 2.39, 10.52 $\pm$ 1.96, 7.01 $\pm$ 1.67) km s$^{-1}$ relative to the LSR derived by \citet{tian2015}.}
\end{deluxetable}

\begin{figure*}
\center\includegraphics[scale=0.51]{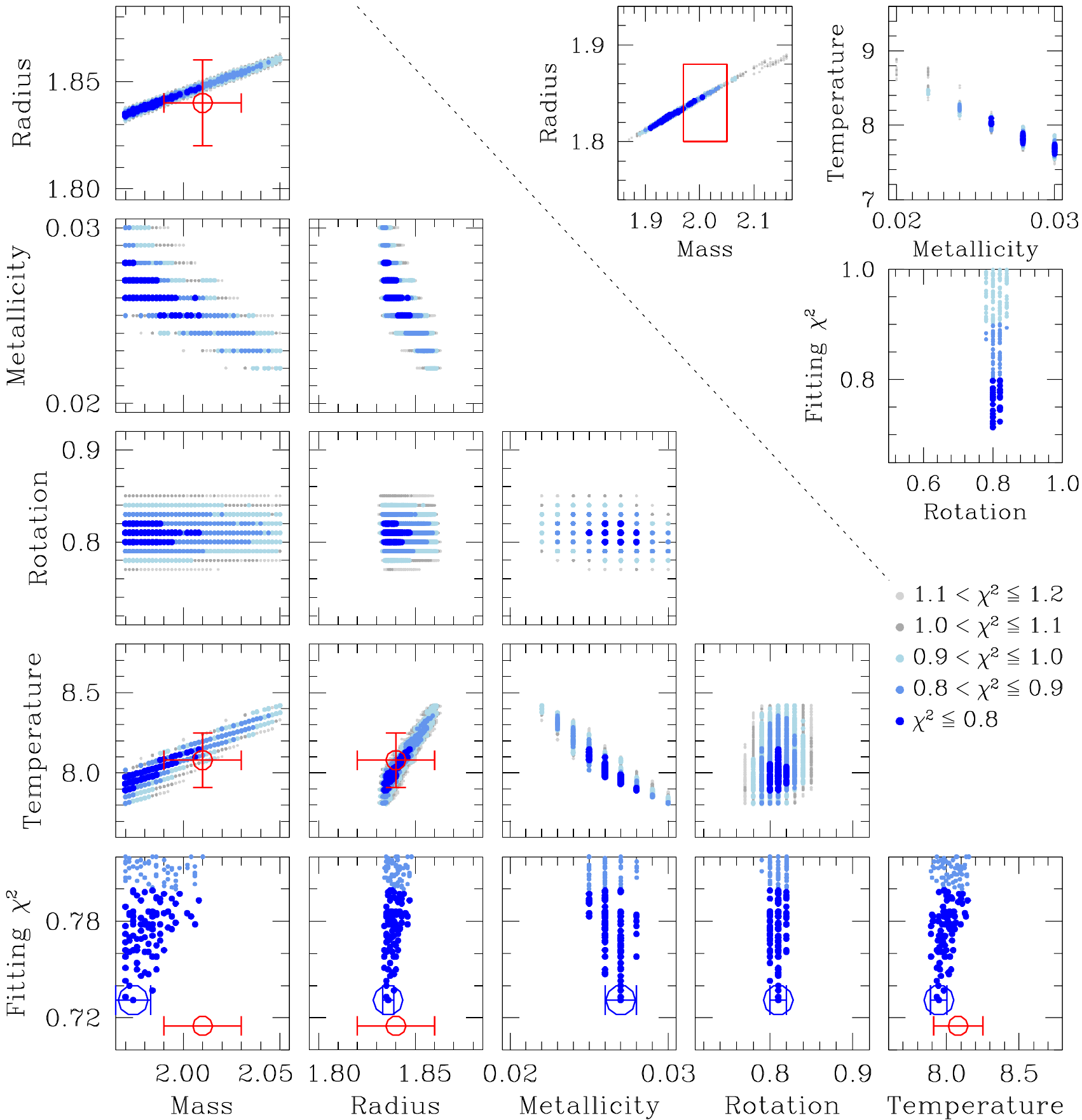}\hskip 3mm\includegraphics[scale=0.51]{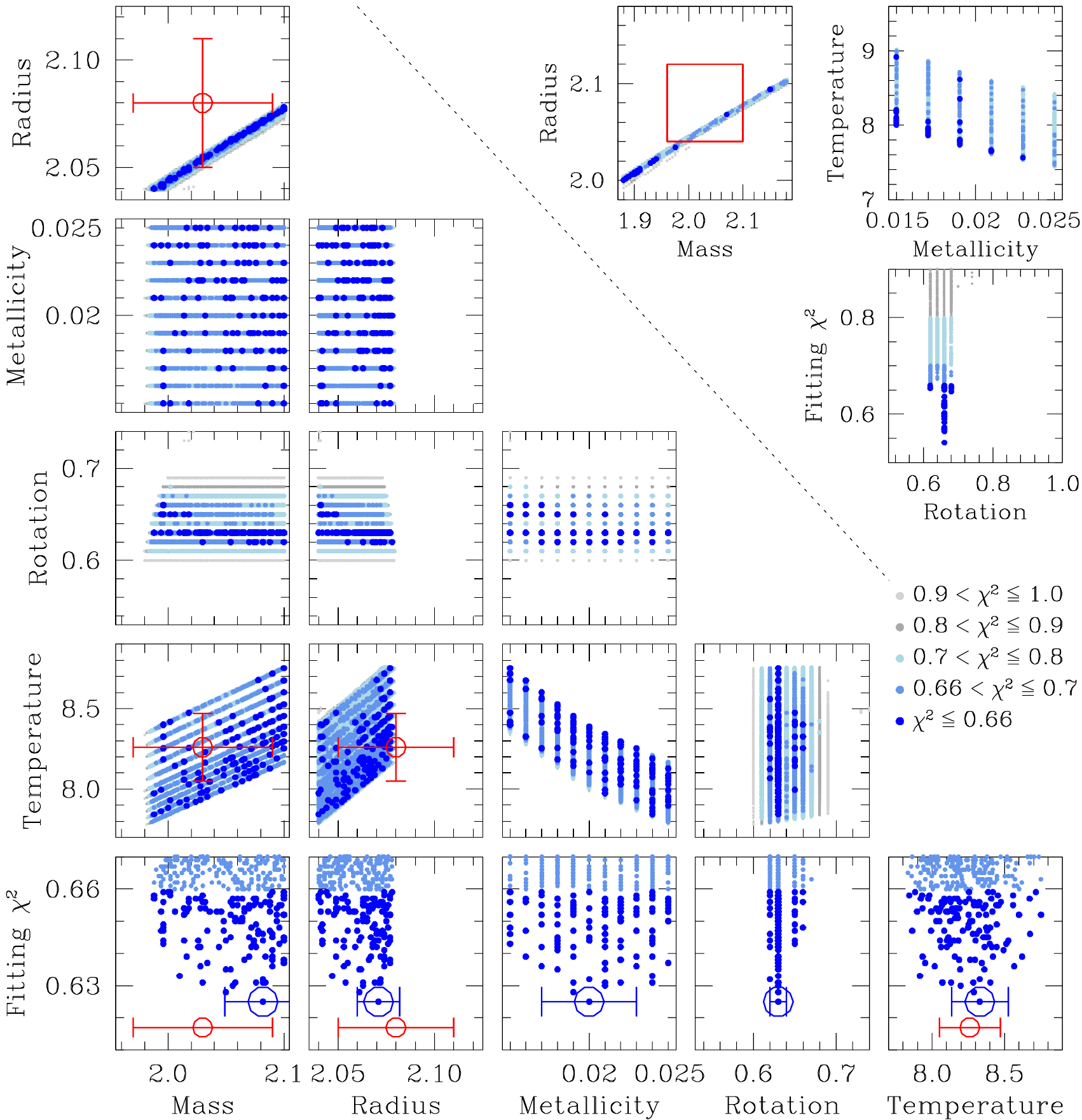}
\caption{Grid fitting plots of AB Cas (left) and OO Dra (right).
The stellar parameters have the units of mass in $M_\sun$, radius in $R_\sun$, metallicity in $Z$, rotation rate in day$^{-1}$, and temperature in 10$^3$ K, respectively.
The three panels on the upper right are for a wide range of stellar masses and radii. The red box corresponds to a narrow range, and its results are shown on the lower-left side.
The red circles with error bars denote the observed values of the mass, radius, and temperature (see Table \ref{tab_param}).
The blue circles with error bars in the bottom panels represent the parameters of the best solution we adopted. \label{fig_Model}}
\end{figure*}

\subsection{Grid-based Fitting \label{grid_fit}}
For a given stellar model, the rotationally shifted theoretical frequencies were fitted to the observed frequencies.
A combination of the theoretical frequencies was determined to obtain a minimum $\chi^2$, defined as:
\begin{equation}
\chi^2 = \frac{1}{N} \sum_{i=1}^{N} \frac{(f_{{\rm theo},i} - f_{{\rm obs},i})^2}{\sigma_i^2},
\end{equation}
where $f_{{\rm theo}, i}$ and $f_{{\rm obs}, i}$ are the theoretical and observed frequencies, respectively, in units of day$^{-1}$.
The uncertainties, $\sigma_i$, were assigned equally to be 0.1 day$^{-1}$, following \citet{murphy2021}, for all frequencies initially.
For the dipole-mode frequency pair with relatively high amplitudes,  $f_{\ell=1, m=1}$ and $f_{\ell=1, m=-1}$ with the same $n$,
the values were later changed to be 0.03 day$^{-1}$ to increase the fitting weight.
Note that the frequency pair is often used to estimate the rotation rate from the well-known equation \citep{goupil2000},
$(f_{m=1} - f_{m=-1})\,/\, 2 \sim f_{\rm rot} \, (C_{n,\ell} - 1)$, where the Ledoux constant $C_{n,\ell}$ is approximately 0.03 for dipole modes with low radial orders of $n \le 5$ \citep{reese2006b}.

Grid-based fitting was performed for various stellar models and rotation rates.
The grids for mass and radius were selected based on the values in Table \ref{tab_param}.
To obtain information on the metallicity of two binary systems, we examined their kinematic properties using Gaia's astrometric data \citep[DR3;][]{Gaia2016, Gaia2022}
and system velocities $\gamma$ in Table  \ref{tab_param}.
The results of Galactic space velocities are listed in Table \ref{tab_kinematic}.
Comparing these results with those of other stars by \citet{chen2021b}, both binaries are likely thin-disk populations. Therefore, metallicities of $Z \sim 0.02$ were considered.

A wide range of masses and radii were first examined. We found several well-fitting models with $\chi^2 < 0.8$ spread out without a definite global minimum,
as shown in the upper-right panels of Figure \ref{fig_Model}.
The range was then narrowed, but the results showed a similar pattern, as presented comprehensively in the lower-left sides of the figure.
It is worth noting that the rotation rates are tightly constrained, although well-fitting models are broadly distributed along the same density lines.
These patterns are identical to our previous study on the eclipsing binary J0247-25 \citep{kim2021}.

\begin{deluxetable*}{rccccc}
\tablewidth{0pt} \tablecolumns{6}
\tablecaption{Comparison of the Observed and Theoretical Frequencies for the Best Solution} \label{tab_seismic}
\tablehead{ \colhead{Observed $f$} & \colhead{Pulsation Mode} & \colhead{Model $f_{n,\ell}$} &  \colhead{Shift $\delta f$} & \colhead{Model $f_{n,\ell,m}$} & \colhead{Difference} }
\startdata
\hline
AB Cas \\
\hline
    $f_1$ = 17.156 & $(1, 0, 0)_{100}$ & 17.313 & $-0.221$ & 17.092 & $+0.064$ \\
    $f_2$ = 18.325 & $(0, 2, 0)_{100}$ & 18.641 & $-0.219$ & 18.422 & $-0.097$ \\
    $f_3$ = 18.244 & $(1, 1, -1)_{100}$ & 17.781 & $+0.468$ & 18.249 & $-0.005$ \\
    $f_4$ = 33.004 & $(4, 1, +1)_{100}$ & 34.654 & $-1.546$ & 33.108 & $-0.104$ \\
$f_{11}$ = 16.678 & $(1, 1, +1)_{100}$ & 17.781 & $-1.105$ & 16.676 & $+0.002$ \\ 
$f_{15}$ = 32.251 & $(3, 3, +1)_{100}$ & 33.415 & $-1.311$ & 32.104 & $+0.147$ \\
$f_{16}$ = 28.290 & $(2, 3, -1)_{100}$ & 27.901 & $+0.317$ & 28.218 & $+0.072$ \\ 
$f_{17}$ = 22.089 & $(2, 0, 0)_{67}$    & 22.385 & $-0.339$ & 22.046 & $+0.043$ \\
$f_{20}$ = 20.044 & $(0, 3, +1)_{100}$ & 20.973 & $-1.004$ & 19.969 & $+0.075$ \\  
$f_{22}$ = 16.887 & $(0, 2, +2)_{100}$ & 18.641 & $-1.714$ & 16.927 & $-0.040$ \\  
$f_{24}$ = 32.280 & $(3, 2, -2)_{100}$ & 31.407 & $+0.778$ & 32.185 & $+0.095$ \\
$f_{31}$ = 37.564 & $(4, 2, -2)_{100}$ & 37.053 & $+0.646$ & 37.699 & $-0.135$ \\  
\hline
OO Dra \\
\hline
    $f_1$ = 41.867 & $(7, 0, 0)_{100}$ & 42.297 & $-0.521$ & 41.776 & $+0.091$ \\
    $f_2$ = 33.473 & $(5, 1, +1)_{100}$ & 34.731 & $-1.234$ & 33.497 & $-0.024$ \\
    $f_3$ = 32.454 & $(4, 2, -2)_{100}$ & 31.928 & $+0.581$ & 32.509 & $-0.055$ \\
    $f_4$ = 34.740 & $(5, 1, -1)_{100}$ & 34.731 & $-0.010$ & 34.721 & $+0.019$ \\
    $f_9$ = 32.631 & $(4, 3, +1)_{100}$ & 33.695 & $-1.047$ & 32.648 & $-0.017$ \\ 
$f_{11}$ = 38.252 & $(6, 1, +1)_{100}$ & 39.686 & $-1.326$ & 38.360 & $-0.108$ \\       
$f_{12}$ = 41.004 & $(6, 3, +3)_{90}$ & 43.626 & $-2.751$ & 40.875 & $+0.129$ \\
$f_{13}$ = 39.519 & $(6, 1, -1)_{100}$ & 39.686 & $-0.097$ & 39.589 & $-0.070$ \\ 
$f_{15}$ = 34.925 & $(5, 2, +2)_{98}$ & 36.905 & $-1.947$ & 34.958 & $-0.033$ \\
$f_{22}$ = 39.616 & $(5, 3, -3)_{90}$ & 38.702 & $+1.023$ & 39.725 & $-0.109$ \\  
$f_{23}$ = 37.547 & $(5, 3, +1)_{100}$ & 38.702 & $-1.129$ & 37.573 & $-0.026$ \\ 
\hline
\enddata
\tablenotetext{}{{\bf Note.} All frequencies have the same unit of day$^{-1}$.}
\end{deluxetable*}

\begin{figure*}
\center\includegraphics[scale=0.64]{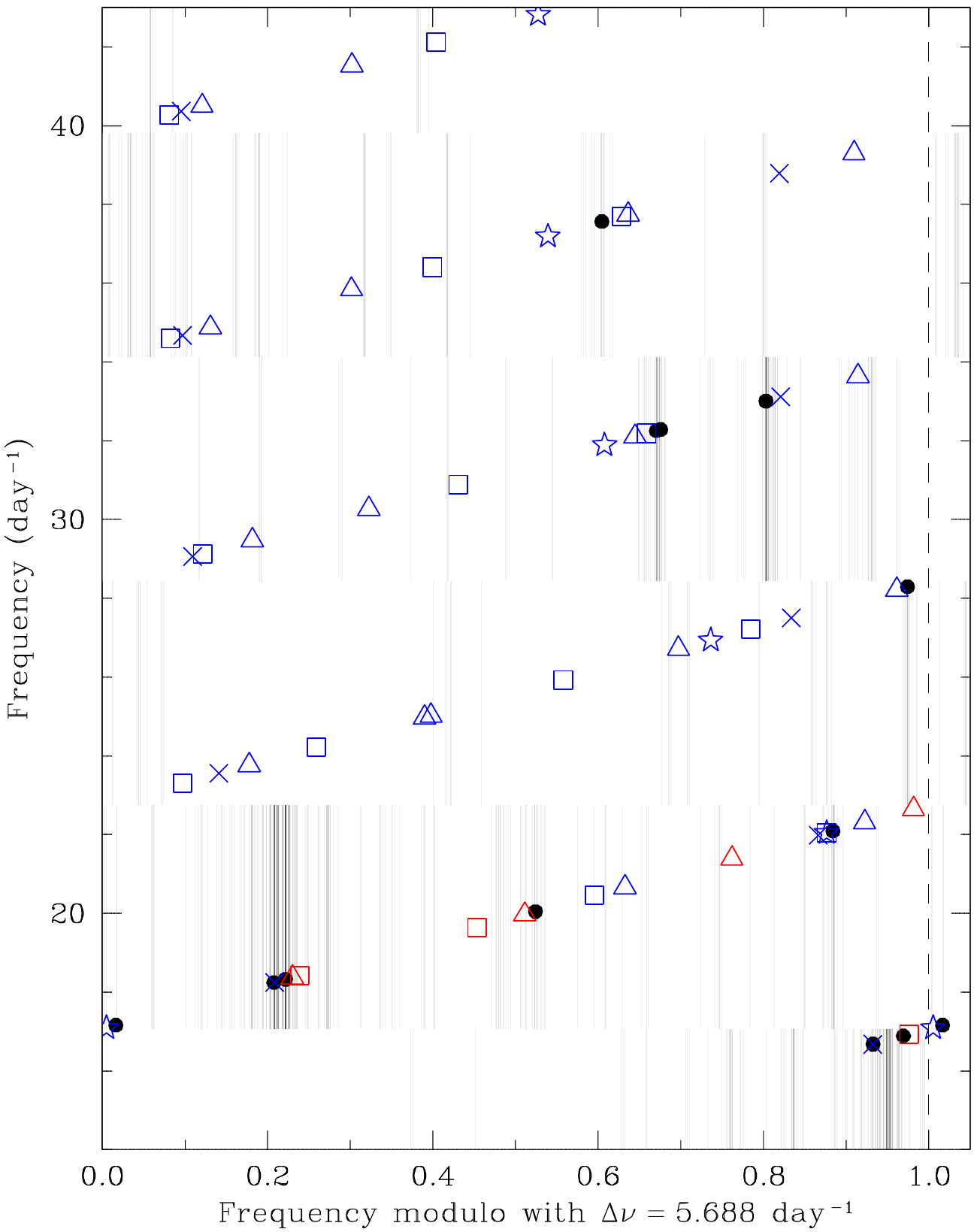}\hskip 5mm\includegraphics[scale=0.64]{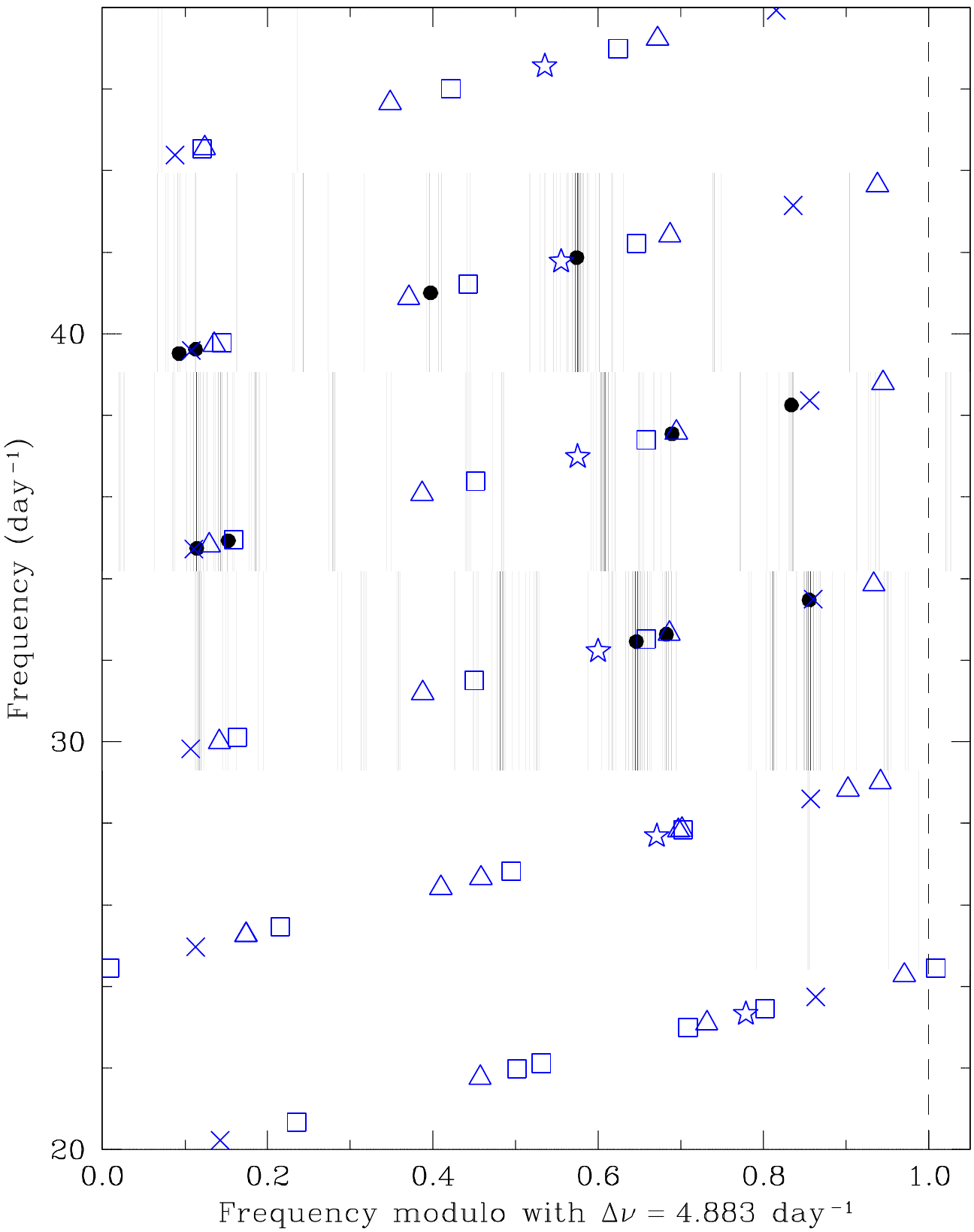}
\caption{\'Echelle diagram of AB Cas (left) and OO Dra (right). The observed frequencies are represented by filled circles. 
The star symbols, crosses, rectangles, and triangles represent the theoretical frequencies, $f_{n,\ell,m}$, with $\ell$ = 0, 1, 2, and 3, respectively.
Red colors on the left denote the nonradial modes with $n$ = 0 and $\ell$ = 2 or 3.
The gray-scale vertical bars represent the Fourier amplitudes of the TESS data; the amplitudes presented on the left were obtained after removing the first frequency of AB Cas. \label{fig_echelle}}
\end{figure*}

\subsection{The Best Solution \label{sec_solution}}
Grid fitting yielded the best solution.
The adopted models with the lowest $\chi^2$ are presented in the bottom panels of Figure \ref{fig_Model}, showing that the theoretical stellar parameters agree with the observations.
The parameters for the AB Cas primary are $M$ = 1.974 $\pm$ 0.009 $M_\sun$, $R$ = 1.836 $\pm$ 0.003 $R_\sun$, $T$ = 7947 $\pm$ 57 K, $Z$ = 0.027 $\pm$ 0.001, and $f_{\rm rot}$ = 0.81 $\pm$ 0.01 day$^{-1}$, 
and those for the OO Dra primary are $M$ = 2.082 $\pm$ 0.033 $M_\sun$, $R$ = 2.071 $\pm$ 0.011 $R_\sun$, $T$ = 8330 $\pm$ 196 K, $Z$ = 0.020 $\pm$ 0.003, and $f_{\rm rot}$ = 0.63 $\pm$ 0.01 day$^{-1}$.
The errors were estimated from standard deviations of 91 well-fitting models with $\chi^2 \le$ 0.80 for AB Cas and those of 136 models with $\chi^2 \le$ 0.66 for OO Dra.

The theoretical frequencies of the best solution are listed in Table \ref{tab_seismic}, which includes the observed frequencies in the first column for comparison.
The second column presents the theoretical pulsation modes $(n, \ell, m)_p$ of the best solution, where $p$ is the distribution rate (\%) among the well-fitting models.
It shows that most well-fitting models have identical pulsation modes to the best solution.
Subsequently, the theoretical frequencies $f_{n,\ell}$ obtained from GYRE for the nonrotating model, their rotational shifts $\delta f$ estimated from the two-dimensional polytropic model,
and the rotationally shifted theoretical frequencies $f_{n,\ell,m}$ = $f_{n,\ell}$ + $\delta f$ follow in order.
The last column shows the differences between the observed and theoretical frequencies.

Figure \ref{fig_echelle} displays the \'echelle diagrams for the best solution, showing that the observed frequencies match the theoretical ones well.
The large frequency separation, $\Delta \nu$ = 5.688 day$^{-1}$ for the AB Cas primary and 4.883 day$^{-1}$ for the OO Dra primary, was derived by averaging the theoretical frequencies with $n$ = 4--6 and $\ell$ = 1 modes.
These values are coincident with those deduced from the empirical relation by \citet{garciahernandez2017}:
5.5 $\pm$ 0.4 day$^{-1}$ for the AB Cas primary with the observed density of 0.323 $\pm$ 0.014 $\rho_\sun$, and 4.6 $\pm$ 0.4 day$^{-1}$ for the OO Dra primary with 0.226 $\pm$ 0.015 $\rho_\sun$.

The pulsation mode is well known only for the dominant frequency $f_1$ of AB Cas, which was identified as a fundamental (F) radial mode ($n$ = 1, $\ell$ = 0)
based on two observational methods: spatial filtration during the eclipse, and photometric amplitude and phase difference between passbands \citep{rodriguez2004a}.
The pulsation constant suggests that $f_1$ of AB Cas is the fundamental radial mode \citep{hong2017}, and $f_1$ of OO Dra is the sixth overtone (6H) radial mode \citep[$n$ = 7, $\ell$ = 0;][]{lee2018}.
The period ratios between radial modes can also be used for mode identification.
The value of $f_1 / f_{17}$ = 0.7767 for AB Cas is slightly higher than the period ratio of P$_{\rm 1H}$ / P$_{\rm F}$ = 0.773 \citep{breger2000},
implying that $f_{17}$ is the first overtone (1H) radial mode ($n$ = 2, $\ell$ = 0).
The slight discrepancy is probably due to the effect of rotation, considering that the period ratio increases with an increase in the star's rotation speed \citep[see Figure 2 by][]{perezhernandez1995}.
These observation-based results agree with the theoretically identified pulsation modes in this study.

\section{Discussion and Conclusion \label{sec_discussion}}
\subsection{Asteroseismic Determination of Rotation Rate}
Following \citet{kim2021}, this is our second study on asteroseismic analysis of fast-rotating $\delta$ Sct stars using the ICC approach,
which estimates the rotational shifts of theoretical pulsation frequencies by fully including the rotation effects based on a two-dimensional polytropic model.
This study revisited two eclipsing binary systems AB Cas and OO Dra, whose primary components are known to be $\delta$ Sct-type pulsators.
By reanalyzing the TESS data, we identified 12 and 11 pulsation frequencies excited in the primary stars of AB Cas and OO Dra, respectively, with amplitudes higher than $\sim$0.3 mmag.
The grid-based seismic analysis yielded the best solution, for which the theoretical frequencies and stellar parameters agreed well with the observations.
These successful seismic results demonstrate the validity of our ICC approach.

It is impressive that the rotation rates were tightly constrained to 0.81 $\pm$ 0.01 day$^{-1}$ for the AB Cas primary and 0.63 $\pm$ 0.01 day$^{-1}$ for the OO Dra primary.
These seismic rotation rates match well with those estimated from the observable properties, 
$v_{\rm eq} / (2\,\pi\,R_{\rm eq})$ = 0.78 $\pm$ 0.04 day$^{-1}$ for the AB Cas primary and 0.68 $\pm$ 0.06 day$^{-1}$ for the OO Dra primary,
where the equatorial rotation velocity $v_{\rm eq}$ is derived from the projected velocity ($v \sin i$) and the orbital inclination ($i$), assuming the rotation axis to align with the orbital axis,
and the equatorial radius $R_{\rm eq}$ is calculated from the observed radius ($R$) and fractional radii ($r_{\rm side}$, $r_{\rm back}$, $r_{\rm volume}$); these parameters are listed in Table \ref{tab_param}.

In contrast to our results, \citet{chen2021a} obtained the seismic rotation rate of the OO Dra primary to be 0.86 $\pm$ 0.01 day$^{-1}$
and concluded that the spin of the primary star is slightly faster than the synchronous value of approximately 0.81 day$^{-1}$.
Their results considerably differ from ours and are inconsistent with the previous spectroscopic observations by \citet{lee2018}.
This discrepancy is probably due to the application of a first-order approximation of rotation effects, $f_{n,\ell,m} = f_{n,\ell} + m\,(1-C_{n,\ell})\,f_{\rm rot}$,  adopted by \citet{chen2021a}.
Since this perturbative approach is valid only for slow rotators, it appears that they misinterpreted the pulsation frequencies of the OO Dra primary.
\citet{miszuda2022} conducted seismic modeling of the AB Cas primary using only the fundamental radial-mode frequency $f_1$, assuming the synchronous rotation ($f_{\rm rot}$ = 0.73159 day$^{-1}$).

\begin{deluxetable}{cccccl}
\tablewidth{0pt} \tablecolumns{5}
\tablecaption{Asteroseismic Rotation Rates of the Primary Components in Close Eclipsing Binaries\label{tab_rot}}
\tablehead{ \colhead{Name} & \colhead{Orbital Frequency} & \colhead{Rotation Rate} & \colhead{$f_{\rm rot}/f_{\rm orb}$} }
\startdata
AB Cas    & 0.731591 day$^{-1}$ & 0.81\,$\pm$\,0.01 day$^{-1}$ & 1.11$^{+0.01}_{-0.02}$ \\
OO Dra    & 0.807506 day$^{-1}$ & 0.63\,$\pm$\,0.01 day$^{-1}$ & 0.78$^{+0.01}_{-0.01}$ \\
J0247-25 & 1.497386 day$^{-1}$ & 1.50\,$\pm$\,0.02 day$^{-1}$ & 1.00$^{+0.02}_{-0.01}$ \\
\enddata
\tablenotetext{}{{\bf Note.} Data of J0247-25 are given by \citet{kim2021}.}
\end{deluxetable}

\subsection{Binary Evolution and Synchronization in Close Binary Systems \label{sec_syncCBS}}
Table \ref{tab_rot} lists three eclipsing binaries for which the rotation rates of the primary components were seismically determined.
These binaries have common characteristics such as short orbital periods of around 1.0 days, low mass ratios of $M_2/M_1$ = 0.09--0.18,
and low masses of the secondary components with 0.19--0.37 $M_\sun$.
Their evolutionary state is different but sequentially linked.
The massive progenitor of the present secondary star evolves earlier into the red giant phase and transfers its mass to the primary component via the Roche-lobe overflow.
AB Cas is a semi-detached Algol that the mass transfer is underway \citep{hong2017} with approximately 5--26 \% loss of the transferred mass \citep{miszuda2022},
whereas OO Dra is in a detached configuration where the mass transfer has been completed \citep{lee2018}.
J0247-25 is an EL CVn-type binary\footnote{The characteristic light curves of EL CVn-type eclipsing binaries show a boxy-shaped primary minimum
due to a total eclipse of a smaller, hotter secondary star by an A-type dwarf primary star \citep{maxted2014,kim2021}.}
of which the small, hot secondary is a highly-evolved star, the remnant of a disrupted red giant and the precursor of an extremely low-mass white dwarf with a helium core of less than 0.3 $M_\sun$ \citep{maxted2014,chen2017,kim2021}.
The AB Cas secondary will contract and heat up after the mass-transfer phase, becoming similar to the OO Dra secondary, 
and then will evolve to a higher temperature at almost constant luminosity like the J0247-25 secondary \citep{hong2017,lee2018}.
Despite the evolutionary change of the secondary star, the mass-enriched primary component remains an A-type dwarf.

In general, stellar components of close binaries are pre-assumed to rotate synchronously with the orbit owing to the strong tidal forces at their small separations.
Theoretical studies show that a binary system tends to evolve toward a state of tidal equilibrium characterized by a circular orbit and synchronous rotation \citep{hut1980,hut1981}.
Observations also indicate that close binaries with orbital periods less than $\sim$5 days are predominantly synchronized \citep{abt2004,dervisoglu2010,lurie2017}.

Most observational studies on rotation-orbit synchronization are based on the projected rotational velocity, $v \sin i$, measured from the spectral line broadening.
It has several sources of measurement error, such as the blending of spectral lines and uncertainty in determining the continuum level \citep{royer2002}.
The projected velocity depends on the rotational inclination, which is usually difficult to measure (see Section \ref{sec_intro});
thus, the orbital inclination is used instead, assuming that the rotation axis aligns with the orbital axis. 
Furthermore, calculating the predicted synchronous velocity $v_{\rm sync}$ to compare with the observed speed requires knowledge of the stellar radius, which seems uncertain.
Another method to deduce the rotation-orbit synchronization is through modelization of the Rossiter-McLaughlin effect observed in radial velocity (RV) data during the eclipse phase.
The modeled RVs are sensitive to the synchronicity parameter (${\sf F}$), defined as the ratio between the rotational and orbital angular velocity 
in the eclipsing binary modeling software PHOEBE \citep[][http://phoebe-project.org/]{prsa2005}.
This approach is less common because it requires a lot of precise RV data during the eclipse.

The semi-detached Algol-type binary RZ Cas is an excellent example studied by these two spectroscopic methods.
According to \citet{lehmann2020}'s careful analysis, the velocity ratio is $v_1 \sin i / v_{1,\rm sync}$ =  0.91 $\pm$ 0.01 and the synchronicity parameter is ${\sf F}_1$ = 0.91 $\pm$ 0.10,
both of which point to the subsynchronous rotation of the primary star; these are average values obtained after excluding the 2001-season data affected probably by the transferred matter.
To calculate $v_{1,\rm sync}$, \citet{lehmann2020} adopted the radius of 1.67 $\pm$ 0.02 $R_\sun$ derived by \citet{rodriguez2004b},
but if the radius of 1.593 $\pm$ 0.007 $R_\sun$ is adopted as in \citet{mkrtichian2018}, the velocity ratio increases to 0.95, getting closer to nearly synchronous rotation.
Moreover, the synchronicity parameter with large dispersion makes it difficult to constrain the synchronization unambiguously.
Another example is the semi-detached Algol, AS Eri.
\citet{lampens2022} obtained the synchronicity parameter of ${\sf F}_1$ = 1.1--1.2 and concluded that the primary star rotates supersynchronously.
It agrees well with the result of rotational velocity that the observed $v_1 \sin i$ = 36 $\pm$ 3 km s$^{-1}$ by \citet{glazunova2008} is a little larger than the predicted $v_{1,\rm sync}$ = 32.9 $\pm$ 0.1 km s$^{-1}$,
adopting the radius of 1.733 $\pm$ 0.006 $R_\sun$ obtained by \citet{lampens2022}.

In contrast to the above spectroscopic methods, we obtained the rotation rates of pulsating primary stars via the asteroseismic analysis, comparing the observed with theoretical pulsation frequencies.
As listed in Table \ref{tab_rot}, our studies show that asteroseismology can be used to precisely measure the rotation rate of fast-rotating $\delta$ Sct stars.
The primary star of J0247-25 is exactly synchronized, with a rotation-to-orbit frequency ratio $f_{\rm rot} / f_{\rm orb}$ = 1.00$^{+0.02}_{-0.01}$.
But the mass-gaining primary component of AB Cas rotates slightly faster than the orbital speed, $f_{\rm rot} / f_{\rm orb}$ = 1.11$^{+0.01}_{-0.02}$, 
probably accelerated by the impact of the gas stream from the secondary donor \citep{lampens2022}.
Conversely, the rotation rate of the OO Dra primary is lower than the synchronous value, $f_{\rm rot} / f_{\rm orb}$ = 0.78 $\pm$ 0.01.
Subsynchronous rotation is uncommon in short-period binaries, and its physical mechanism is not yet well understood \citep{torres2010,lehmann2020}.
Although based on limited samples, our results for the three binaries imply that more binaries than widely expected have stellar components to rotate asynchronously with the orbital motion.
Therefore, the assumption of synchronous rotation should be taken with some caution.
Further asteroseismic samples are required to provide a definitive constraint on the rotation-orbit synchronization in close binary systems.

\begin{acknowledgements}
We are grateful to an anonymous referee for valuable comments and suggestions. We wish to thank Dr. J. W. Lee for his helpful comments.
This research was supported by the Korea Astronomy and Space Science Institute under the R\&D program (Project No. 2023-1-832-03) supervised by the Ministry of Science and ICT.
This paper includes data collected by the TESS mission. Funding for the TESS mission is provided by the NASA's Science Mission Directorate.
This work has made use of data from the European Space Agency (ESA) mission {\it Gaia} (\url{https://www.cosmos.esa.int/gaia}), processed by the {\it Gaia}
Data Processing and Analysis Consortium (DPAC, \url{https://www.cosmos.esa.int/web/gaia/dpac/consortium}).
Funding for the DPAC has been provided by national institutions, in particular, the institutions participating in the {\it Gaia} Multilateral Agreement.
We thank Editage (www.editage.com) for English language editing.
\end{acknowledgements}

\facility{TESS \citep{ricker2015}}

\appendix
\section{List of observed frequencies}

\begin{deluxetable*}{lrlclrlcr}
\tabletypesize{\footnotesize}
\tablenum{A1}
\tablecaption{Multiple Frequencies of AB Cas Detected from the TESS Data\label{tab_ABobsfreq} }
\tablehead{
\multicolumn{3}{c}{This Work} & & \multicolumn{3}{c}{\citet{miszuda2022}} & &\multicolumn{1}{r}{Difference} \\
\cline{1-3} \cline{5-7}
\colhead{Frequency} & \colhead{Amplitude} & \colhead{Remark}  & & \colhead{Frequency} & \colhead{Amplitude} & \colhead{Remark} & & $f_i - f_{{\rm M22}\_j}$ }
\startdata
\multicolumn{9}{l}{\it Frequencies Detected Identically in Two Results} \\
$\,\:f_1$ = {\bf 17.15643(0)}  & 14.60(2) & {\bf $\delta$ Sct-type} & & $\,\:f_{\rm M22\_1}$ = {\bf 17.15643(0)}  & 12.30(1) & {\bf Independent} & & $0.00000$ \\
$\,\:f_2$ = {\bf 18.32462(2)}  & 1.37(2)   & {\bf $\delta$ Sct-type} & & $\,\:f_{\rm M22\_2}$ = {\bf 18.32466(3)}  & 1.20(1)   & {\bf Independent} & & $-0.00004$ \\
$\,\:f_3$ = {\bf 18.24374(3)}  & 1.30(2)   & {\bf $\delta$ Sct-type} & & $\,\:f_{\rm M22\_3}$ = {\bf 18.24361(3)}  & 1.20(1)   & {\bf Independent} & & $+0.00013$ \\
$\,\:f_4$ = {\bf 33.00379(2)}  & 1.18(2)   & {\bf $\delta$ Sct-type} & & $\,\:f_{\rm M22\_4}$ = {\bf 33.00379(4)}  & 1.02(1)   & {\bf Independent} & & $0.00000$ \\
$\,\:f_5$ = \,16.78048(3)       & 0.99(2)   & $f_3 - 2\,f_{\rm orb}$         & & $\,\:f_{\rm M22\_5}$ = \,16.78036(4)       & 0.99(1)   & $f_{\rm M22\_3} - 2\,f_{\rm orb}$ & & $+0.00012$ \\
$\,\:f_6$ = \,34.46699(3) 	   & 0.78(2)   & $f_4 + 2\,f_{\rm orb}$         & & $f_{\rm M22\_17} $ = \,34.46691(5)       & 0.63(1)   & $f_{\rm M22\_4} + 2\,f_{\rm orb}$ & & $+0.00008$ \\
$\,\:f_7$ = \,\, 0.06142(5)      & 0.71(2)   &  -                                  & & $\,\:f_{\rm M22\_7}$ = {\bf \, 0.06586(5)}   & 0.76(1)   & {\bf Independent} & & $-0.00444$    \\
$\,\:f_8$ = \,34.31282(4) 	   & 0.67(2)   & $2\,f_1$                       & & $f_{\rm M22\_13}$ = \,34.31280(5) 	   & 0.62(1)   & $2\,f_{\rm M22\_1}$  & & $+0.00002$  \\
$\,\:f_9$ = \,15.69325(4) 	   & 0.73(2)   & $f_1 - 2\,f_{\rm orb}$         & & $f_{\rm M22\_15}$ = \,15.69316(5) 	   & 0.62(1)   & $f_{\rm M22\_1} - 2\,f_{\rm orb}$ & & $+0.00009$  \\
$f_{10}$ = \,18.61949(4) 	   & 0.66(2)   & $f_1 + 2\,f_{\rm orb}$        & & $f_{\rm M22\_14}$ = \,18.61990(5) 	   & 0.64(1)   & $f_{\rm M22\_1} + 2\,f_{\rm orb}$ & & $-0.00041$   \\ 
$f_{11}$ = {\bf 16.67814(5)} & 0.64(2)   & {\bf $\delta$ Sct-type}  & & $f_{\rm M22\_19}$ = {\bf 16.67837(5)} & 0.51(1)   & {\bf Independent} & & $-0.00023$ \\
$f_{12}$ = \,18.25800(7)      & 0.61(2)   & $f_2 - f_7$                   & & $f_{\rm M22\_22}$ = \,18.25825(6)       & 0.51(1)   & $f_{\rm M22\_2} - f_{\rm M22\_7}$ & & $-0.00025$ \\
$f_{15}$ = {\bf 32.25056(7)} & 0.42(2)   & {\bf $\delta$ Sct-type} & & $f_{\rm M22\_34}$ = \, 32.25047(8)      & 0.31(1)   &$2\,f_{\rm M22\_4} - 2\,f_{\rm M22\_32}$ & & $+0.00009$ \\
$f_{16}$ = {\bf 28.28972(6)} & 0.47(1)   & {\bf $\delta$ Sct-type} & & $f_{\rm M22\_24}$ = {\bf 28.28968(7)}  & 0.40(1)   & {\bf Independent} & & $+0.00004$ \\
$f_{17}$ = {\bf 22.08932(7)} & 0.45(1)   & {\bf $\delta$ Sct-type} & & $f_{\rm M22\_25}$ = {\bf 22.08924(7)}  & 0.40(1)   & {\bf Independent} & & $+0.00008$ \\                      
$f_{19}$ = \,16.12973(7) 	   & 0.43(2)   & $f_2 - 3\,f_{\rm orb}$         & & $f_{\rm M22\_27}$ = \,16.12969(7) 	   & 0.40(1)   & $f_{\rm M22\_2} - 3\,f_{\rm orb}$ & & $+0.00004$ \\
$f_{20}$ = {\bf 20.04429(7)} & 0.41(1)   & {\bf $\delta$ Sct-type} & & $f_{\rm M22\_31}$ = {\bf 20.04429(8)}  & 0.32(1)   & {\bf Independent} & & $0.00000$  \\    
$f_{22}$ =  {\bf 16.88736(7)} & 0.39(1)  & {\bf $\delta$ Sct-type} & & $f_{\rm M22\_32}$ = \,16.88179(8)        & 0.32(1)  & $3\,f_{\rm M22\_7} - f_{\rm M22\_19}$ & & $+0.00557$  \\
$f_{23}$ = \,19.78787(7)   	    & 0.39(2)   & $f_2 + 2\,f_{\rm orb}$       & & $f_{\rm M22\_26}$ = \,19.78790(7)   	    & 0.38(1)   & $f_{\rm M22\_2} + 2\,f_{\rm orb}$ & & $-0.00003$ \\
$f_{24}$ = {\bf 32.28046(7)} & 0.41(2)   & {\bf $\delta$ Sct-type} & & $f_{\rm M22\_23}$ = {\bf 32.28075(6)}   & 0.36(1)   & {\bf Independent} & & $-0.00029$ \\ 				
$f_{28}$ = \,36.49344(8) 	   & 0.35(1)   & $2\,f_3$                      & & $f_{\rm M22\_36}$ = \,36.49343(9) 	   & 0.30(1)   & $2\,f_{\rm M22\_3}$ & & $+0.00001$ \\			
$f_{30}$ = \,40.14989(9) 	   & 0.34(1)   & $2\,f_3 + 5\,f_{\rm orb}$    & & $f_{\rm M22\_39}$ = \,40.14994(9) 	   & 0.28(1)   & $2\,f_{\rm M22\_3} + 5\,f_{\rm orb}$ & & $-0.00005$ \\	
$f_{31}$ = {\bf 37.56356(9)} & 0.33(1)   & {\bf $\delta$ Sct-type} & & $f_{\rm M22\_33}$ = {\bf 37.55835(8)} & 0.32(1)   & {\bf Independent} & & $+0.00521$ \\		
$f_{32}$ =	 \,18.14140(9) 	   & 0.31(2)   & $f_{11} + 2\,f_{\rm orb}$ 	 & & $f_{\rm M22\_59}$ =	\,18.14140(12) 	  & 0.19(1)   & $f_{\rm M22\_19} + 2\,f_{\rm orb}$ & & $0.00000$ \\
\\
\multicolumn{9}{l}{\it Frequencies Detected in Only One Result or Frequencies of Which Amplitudes Are Significantly Different between the Two Results} \\
$f_{13}$ = \,\, 1.03275(5) 	   & 0.57(1)   & -                                   & & \multicolumn{3}{c}{-} & & - \\
$f_{14}$ = \,\, 1.32253(6) 	   & 0.51(2)   & -                                   & & \multicolumn{3}{c}{-} & & - \\
$f_{18}$ = \,\, 0.08316(6)     & 0.61(2)   & $f_2 - f_3 $	          & & \multicolumn{3}{c}{-} & & - \\
$f_{21}$ = \,\, 1.62145(15)   & 0.38(2)   & -                                   & & \multicolumn{3}{c}{-} & & - \\ 
$f_{25}$ = \,\, 1.44926(10)   & 0.38(2)   & -                         	          & & \multicolumn{3}{c}{-} & & - \\
$f_{26}$ = \,\, 0.12698(7) 	   & 0.50(2)   & $2\,f_7$                       & & \multicolumn{3}{c}{-} & & - \\
$f_{27}$ = \,\, 0.20347(8) 	   & 0.38(2)   & $f_{22} - f_{11}$	          & & \multicolumn{3}{c}{-} & & - \\
$f_{29}$ =	 \,\, 1.20703(8) 	   & 0.35(1)   & -                                  & & \multicolumn{3}{c}{-} & & - \\
\multicolumn{3}{c}{-}                                                                     & & $\,\:f_{\rm M22\_6}$ = \,17.88798(4)       & 0.83(1)   & $f_{\rm M22\_1} + f_{\rm orb}$ & & - \\
\multicolumn{3}{c}{-}                                                                     & & $\,\:f_{\rm M22\_8}$ = \,16.42488(4)       & 0.78(1)   & $f_{\rm M22\_1} - f_{\rm orb}$ & & - \\
\multicolumn{3}{c}{-}                                                                     & & $\,\:f_{\rm M22\_9}$ = \,14.23006(4)       & 0.74(1)   & $f_{\rm M22\_1} - 4\,f_{\rm orb}$ & & - \\
$f_{157}$ = 19.35167(23) 	   & 0.12(1)   & $f_1 + 3\,f_{\rm orb}$       & & $f_{\rm M22\_10}$ = \,19.35128(4)         & 0.74(1)   & $f_{\rm M22\_1} + 3\,f_{\rm orb}$ & & $+0.00039$ \\
\multicolumn{3}{c}{-}                                                                     & & $f_{\rm M22\_11}$ = \,20.08278(4)        & 0.75(1)   & $f_{\rm M22\_1} + 4\,f_{\rm orb}$ & & - \\
$f_{113}$ = 14.96132(20) 	   & 0.12(1)   & $f_1 - 3\,f_{\rm orb}$        & & $f_{\rm M22\_12}$ = \,14.96156(4)         & 0.72(1)   & $f_{\rm M22\_1} - 3\,f_{\rm orb}$ & & $-0.00024$ \\
$\,\:f_{94}$ = 13.49851(14)  & 0.15(1)   & $f_1 - 5\,f_{\rm orb}$        & & $f_{\rm M22\_16}$ = \,13.49851(5)         & 0.57(1)   & $f_{\rm M22\_1} - 5\,f_{\rm orb}$ & & $0.00000$ \\
$f_{158}$ = 20.81498(21) 	   & 0.10(1)   & $f_1 + 5\,f_{\rm orb}$       & & $f_{\rm M22\_18}$ = \,20.81438(5)         & 0.55(1)   & $f_{\rm M22\_1} + 5\,f_{\rm orb}$ & & $+0.00060$ \\
\multicolumn{3}{c}{-}                                                                     & & $f_{\rm M22\_20}$ = \,21.54598(5)        & 0.50(1)   & $f_{\rm M22\_1} + 6\,f_{\rm orb}$ & & - \\
\multicolumn{3}{c}{-}                                                                     & & $f_{\rm M22\_21}$ = \,12.76696(6)        & 0.49(1)   & $f_{\rm M22\_1} - 6\,f_{\rm orb}$ & & - \\
\multicolumn{3}{c}{-}                                                                     & & $f_{\rm M22\_28}$ = \,22.27734(7)        & 0.36(1)   & $f_{\rm M22\_1} + 7\,f_{\rm orb}$ & & - \\
\multicolumn{3}{c}{-}                                                                     & & $f_{\rm M22\_29}$ = \,12.03528(8)        & 0.34(1)   & $f_{\rm M22\_1} - 7\,f_{\rm orb}$ & & - \\
\multicolumn{3}{c}{-}                                                                     & & $f_{\rm M22\_30}$ = \,17.55066(8)        & 0.33(1)   & $f_{\rm M22\_1} + 6\,f_{\rm M22\_7}$ & & - \\
\multicolumn{3}{c}{-}                                                                     & & $f_{\rm M22\_35}$ = \,11.30366(8)        & 0.30(1)   & $f_{\rm M22\_1} - 8\,f_{\rm orb}$ & & - \\
\multicolumn{3}{c}{-}                                                                     & & $f_{\rm M22\_37}$ = \,16.83426(8)        & 0.32(1)   & $-5\,f_{\rm M22\_19} + 5\,f_{\rm M22\_31}$ & & - \\
\enddata
\tablenotetext{}{{\bf Note.} The frequencies are in units of day$^{-1}$, and the amplitudes have units of mmag in this work and ppt in \citet{miszuda2022}; 1.0 ppt is equal to 1.0863 mmag.
The values in parentheses are the errors in the last digits.}
\end{deluxetable*}

\begin{deluxetable*}{lrlclrlcr}
\tabletypesize{\footnotesize}
\tablenum{A2}
\tablecaption{Multiple Frequencies of OO Dra Detected from the TESS Data\label{tab_OOobsfreq} }
\tablehead{
\multicolumn{3}{c}{This Work} & & \multicolumn{3}{c}{\citet{chen2021a}} & &\multicolumn{1}{r}{Difference} \\
\cline{1-3} \cline{5-7}
\colhead{Frequency} & \colhead{Amplitude} & \colhead{Remark}  & & \colhead{Frequency} & \colhead{Amplitude} & \colhead{Remark} & & $f_i - f_{{\rm C21}\_j}$ }
\startdata
\multicolumn{9}{l}{\it Frequencies Detected Identically in Two Results} \\
$\,\:f_1$ = {\bf 41.86670(1)}  & 2.25(1) \, & {\bf $\delta$ Sct-type} & & $\,\:f_{\rm C21\_1}$ = {\bf 41.8669(4)}  & 2.51(2) & {\bf Independent} & & $-0.0002$ \\
$\,\:f_2$ = {\bf 33.47314(8)}  & 2.19(59)  & {\bf $\delta$ Sct-type} & & $\,\:f_{\rm C21\_2}$ = {\bf 33.4728(5)}  & 2.21(2) & {\bf Independent} & & $+0.0003$ \\
$\,\:f_3$ = {\bf 32.45409(0)}  & 1.77(1) \, & {\bf $\delta$ Sct-type} & & $\,\:f_{\rm C21\_4}$ = {\bf 32.4538(7)}  & 1.63(2) & {\bf Independent} & & $+0.0003$ \\
$\,\:f_4$ = {\bf 34.74007(0)}  & 1.83(1) \, & {\bf $\delta$ Sct-type} & & $\,\:f_{\rm C21\_3}$ = {\bf 34.7407(6)}  & 1.94(2) & {\bf Independent} & & $-0.0006$ \\
$\,\:f_5$ = \,31.65693(1)       & 1.19(1) \, & $f_3 - f_{\rm orb}$             & & $\,\:f_{\rm C21\_5}$ = \,31.6572(11)      & 1.02(2) & $f_{\rm C21\_4} - f_{\rm orb}$ &  & $-0.0003$ \\
$\,\:f_8$ = \,33.26365(1)       & 0.74(1) \, & $f_3 + f_{\rm orb}$            & & $\,\:f_{\rm C21\_6}$ = \,33.2641(17)      & 0.69(2) & $f_{\rm C21\_4} + f_{\rm orb}$ & & $-0.0004$ \\
$\,\:f_9$ = {\bf 32.63133(1)}  & 0.69(1) \, & {\bf $\delta$ Sct-type} & & $\,\:f_{\rm C21\_7}$ = {\bf 32.6314(20)} & 0.58(2) & {\bf Independent} & & $-0.0001$ \\
$f_{10}$ = \,33.12505(1)       & 0.51(1) \, & $f_4 - 2\,f_{\rm orb}$         & & $\,\:f_{\rm C21\_9}$ = \,33.1256(26)       & 0.45(2) & $f_{\rm C21\_3} - 2\,f_{\rm orb}$ & & $-0.0005$ \\ 
$f_{11}$ =  {\bf 38.25249(1)} & 0.49(1) \, & {\bf $\delta$ Sct-type} & & $f_{\rm C21\_11}$ = {\bf 38.2527(31)}    & 0.38(2) & {\bf Independent} & & $-0.0002$ \\
$f_{12}$ = {\bf 41.00371(2)} & 0.41(1) \, & {\bf $\delta$ Sct-type} & & $f_{\rm C21\_12}$ = {\bf 41.0008(35)}    & 0.38(2) & {\bf Independent} & & $+0.0029$ \\
$f_{13}$ = {\bf 39.51906(2)} & 0.34(1) \, & {\bf $\delta$ Sct-type} & & $f_{\rm C21\_23}$ = \,39.5176(55)          & 0.25(2) &$f_{\rm C21\_3} + f_{\rm C21\_11} - f_{\rm C21\_2}$ & & $+0.0015$ \\
$f_{14}$ = \,34.87859(2)       & 0.37(1) \, & $f_3 + 3\,f_{\rm orb}$       & & $f_{\rm C21\_10}$ = \,34.8789(29)          & 0.41(2) & $f_{\rm C21\_4} + 3\,f_{\rm orb}$ & & $-0.0003$ \\
$f_{15}$ = {\bf 34.92535(2)} & 0.32(1) \, & {\bf $\delta$ Sct-type} & & $f_{\rm C21\_17}$ = \,34.9245(37)          & 0.31(2) & $f_{\rm C21\_3} + f_{\rm C21\_7} - f_{\rm C21\_4}$ & & $+0.0009$ \\
$f_{17}$ = \,28.58489(2)       & 0.34(1) \, & $f_9 - 5\,f_{\rm orb}$        & & $f_{\rm C21\_16}$ = \,28.5868(33)          & 0.35(2) & $f_{\rm C21\_7} - 5\,f_{\rm orb}$ & & $-0.0019$ \\                      
$f_{18}$ = \,35.08796(2)       & 0.34(1) \, & $f_2 + 2\,f_{\rm orb}$       & & $\,\:f_{\rm C21\_8}$ = \,35.0878(25)        & 0.46(2) & $f_{\rm C21\_2} + 2\,f_{\rm orb}$ & & $+0.0002$ \\
$f_{19}$ = \,30.04188(2)       & 0.32(1) \, & $f_3 - 3\,f_{\rm orb}$        & & $f_{\rm C21\_19}$ = \,30.0423(43)          & 0.27(2) & $f_{\rm C21\_4} - 3\,f_{\rm orb}$ & & $-0.0004$ \\
$f_{20}$ = \,\, 0.79475(3)      & 0.30(1) \, & $f_{\rm orb}$                    & & $f_{\rm C21\_14}$ = \,\, 0.7938(33)         & 0.36(2) & $f_{\rm orb}$ & & $+0.0010$ \\ 
$f_{22}$ = {\bf 39.61602(2)} & 0.29(1) \, & {\bf $\delta$ Sct-type} & & $f_{\rm C21\_22}$ = \,39.6148(42)          & 0.26(2) & $f_{\rm C21\_1} + f_{\rm C21\_7} - f_{\rm C21\_10}$ & & $+0.0012$ \\
$f_{23}$ = {\bf 37.54728(2)} & 0.29(1) \, & {\bf $\delta$ Sct-type} & & $f_{\rm C21\_21}$ = \,37.5477(43)          & 0.28(2) & $2\,f_{\rm C21\_8} - f_{\rm C21\_7}$ & & $-0.0004$ \\
$f_{25}$ = \,36.53852(3)      & 0.28(1) \, & $f_{15} + 2\,f_{\rm orb}$    & & $f_{\rm C21\_18}$ = \,36.5401(41)          & 0.29(2) & $f_{\rm C21\_17} + 2\,f_{\rm orb}$ & & $-0.0016$ \\	
$f_{26}$ = \,39.38451(3)      & 0.26(1) \, & $f_{12} - 2\,f_{\rm orb}$    & & $f_{\rm C21\_13}$ = \,39.3842(31)          & 0.37(2) & $f_{\rm C21\_12} - 2\,f_{\rm orb}$ & & $+0.0003$ \\	
\\
\multicolumn{9}{l}{\it Frequencies Detected in Only One Result or Frequencies of Which Amplitudes Are Significantly Different between the Two Results} \\
$\,\:f_6$ = \,37.14252(1)       & 1.11(1) \, & $46\,f_{\rm orb}$               & & $f_{\rm C21\_20}$ = \,37.1334(46)         & 0.27(2) & $f_{\rm C21\_15} + 9\,f_{\rm orb}$ & & $+0.0091$ \\	
$\,\:f_7$ = \,29.87309(1)       & 0.76(1) \, & $37\,f_{\rm orb}$               & & $f_{\rm C21\_15}$ = \,29.8652(37)         & 0.36(2) & $f_{\rm C21\_2} + f_{\rm C21\_11} - f_{\rm C21\_1}$ & &  $+0.0079$ \\
$f_{16}$ = \,33.47364(19)     & 0.90(60)  & $f_2$                          & & \multicolumn{3}{c}{-} & & - \\
$f_{21}$ = \,\, 0.07133(2)      & 0.29(1) \, & -                                  & & \multicolumn{3}{c}{-} & & - \\
$f_{24}$ = \,41.86564(22)     & 0.30(2) \, & $f_1$                          & & \multicolumn{3}{c}{-} & & - \\
\enddata
\tablenotetext{}{{\bf Note.} The frequencies are in units of day$^{-1}$, and the amplitudes have units of mmag. The values in parentheses are the errors in the last digits.}
\end{deluxetable*}

\end{document}